\documentclass{emulateapj}
\usepackage{apjfonts}
\usepackage{graphicx}

\usepackage[usenames,dvipsnames]{color}
\usepackage{amssymb,amsmath}
\usepackage{url}

\slugcomment{ApJ: Accepted 14 Oct 2013}
\shorttitle{\sc Velocity shifts and distortions in optical spectra}
\shortauthors{\sc Evans \& Murphy}

\begin{document}


\title{A new method for detecting velocity shifts and distortions between optical spectra}


\author{\sc
Tyler M. Evans\altaffilmark{1}
and
Michael T. Murphy\altaffilmark{1}
}
                                                                                
\altaffiltext{1}{Centre for Astrophysics and Supercomputing, Swinburne University of Technology, Hawthorn, Victoria 3122,
Australia {\tt tevans@astro.swin.edu.au}}

\begin{abstract}
Recent quasar spectroscopy from the VLT and Keck telescopes suggests that fundamental constants may not actually be constant. To better confirm or refute this result, systematic errors between telescopes must be minimized.  We present a new method to directly compare spectra of the same object and measure any velocity shifts between them.  This method allows for the discovery of wavelength-dependent velocity shifts between spectra, i.e.~velocity distortions, that could produce spurious detections of cosmological variations in fundamental constants.  This ``direct comparison'' method has several advantages over alternative techniques: it is model-independent (cf.~line-fitting approaches), blind, in that spectral features do not need to be identified beforehand, and it produces meaningful uncertainty estimates for the velocity shift measurements.  In particular, we demonstrate that, when comparing echelle-resolution spectra with unresolved absorption features, the uncertainty estimates are reliable for signal-to-noise ratios $\gtrsim$7\,per pixel.  We apply this method to spectra of quasar J2123$-$0050 observed with Keck and the VLT and find no significant distortions over long wavelength ranges ($\sim1050$\,\AA) greater than $\approx$180\,m\,s$^{-1}$. We also find no evidence for systematic velocity distortions within echelle orders greater than 500\,m\,s$^{-1}$. Moreover, previous constraints on cosmological variations in the proton--electron mass ratio should not have been affected by velocity distortions in these spectra by more than $4.0\pm4.2$\, parts per million. This technique may also find application in measuring stellar radial velocities in search of extra-solar planets and attempts to directly observe the expansion history of the Universe using quasar absorption spectra.
\end{abstract}




\section{Introduction}
\label{sec:DC_intro}
The Standard Model of particle physics is one of the most successful theories of modern physics but, at the same time, it embodies a major area of ignorance. Within it, there are several values referred to as `fundamental constants', which contain no underlying physical understanding -- they must be measured in the laboratory and inserted into the Model without knowing why or where they come from.  Two such values are the fine-structure constant,  $\alpha \equiv e^{2}/\hbar c$, and the ratio of the proton and electron masses, $\mu \equiv m_{\rm p}/m_{\rm e}$.  The implication of varying `fundamental constants' demonstrates how the Standard Model may be incomplete: if fundamental constants actually vary with time or space, a more fundamental model of particle physics will likely  be required. For example, various string theories have hypothesized varying fundamental constants that couple to extra, compactified dimensions (e.g.~\citealt{Damour:1994:532}).  Furthermore, if a `grand unified theory' is eventually successful, it may give some explanation for the values the fundamental constants take, as well as how they depend on other parameters in the new theory \citep[e.g.][]{Uzan:2003:403}.

Starting with \citet{Webb:1999:884}, there has been tentative evidence that the $\alpha$ may actually be variable.  In contrast, all measurements of $\mu$ are consistent with no variation at a precision level of a few $\times 10^{-7}$ (e.g.~\citealt{Bagdonaite:2013:46}) at redshifts $z<$1 and $\sim$10$^{-6}$ at $z>2$ \citep[e.g.][]{Weerdenburg:2011:180802}.  These fundamental constants are measured at different times and places and compared to their Earth-based, laboratory values. The measured quantity is the relative difference between the value of $\alpha$ on cosmic scales and the laboratory value:
\begin{equation}
\Delta \alpha/\alpha \equiv \frac{\alpha_{\rm obs}-\alpha_{\rm lab}}{\alpha_{\rm lab}} \approx \frac{-\Delta v}{2c\Delta Q},
\label{eq:alpha}
\end{equation}
where, for the case of two spectral lines, $\Delta v$ is the velocity shift between the lines caused by a varying $\alpha$, $c$ is the speed of light and $\Delta Q$ is defined as $q_{2}/\omega_{2}-q_{1}/\omega_{1}$ where $q_{i}$ is a measure of the sensitivity of line $i$ to variations in $\alpha$, and $\omega_i$ is the wavenumber of the transition.  Likewise, $\mu$ is related to velocity shifts between transitions via the following equation:
\begin{equation}
\Delta \mu/\mu \equiv \frac{\mu_{\rm obs}-\mu_{\rm lab}}{\mu_{\rm lab}} \approx \frac{\Delta v}{cK}. 
\label{eq:mu}
\end{equation}
Here, $K$ determines the sensitivity of a transition to the value of $\mu$.

To measure  $\Delta \alpha/\alpha$, \citet{Webb:1999:884} applied equation 1 to many transitions from many different ionic metal species simultaneously -- the Many Multiplet method \citep{Dzuba:1999:230} -- to 30 quasar (QSO) absorption systems in the range  $0.5<z<1.6$.  In a series of following papers, the number of  Keck/HIRES-observed absorption systems was increased to $143$, giving a weighted mean value of $\Delta \alpha/\alpha=(-5.7\pm 1.1)\times 10^{-6}$ and covering a redshift range of $0.2<z<4.2$ \citep{Murphy:2001:1208, Murphy:2003:609, Murphy:2004:131}. A similar technique is used to measure $\Delta \mu/\mu$, however the primary constraints for $\mu$ come from the molecular hydrogen (e.g.~\citealt{Thompson:1975:3, Malec:2010:1541}), in rare cases ammonia (e.g.~\citealt{Flambaum:2007:240801,Murphy:2008:1611,Kanekar:2011:L12}) and, more recently, methanol (e.g.~\citealt{Ellingsen:2012:L7,Bagdonaite:2013:46}).  Therefore, while the techniques to measure $\Delta \alpha/\alpha$ and $\Delta \mu/\mu$ are similar, for $\alpha$ the relevant probe in optical spectra is metal ion transitions, and for $\mu$ it is Lyman and Werner molecular hydrogen transitions falling in the Lyman-$\alpha$ forest region of quasar spectra.

In the years following \citet{Murphy:2004:131}, a few small samples of absorption systems were published (e.g.~\citealt{Chand:2004:853}, cf.~\citealt{Murphy:2008:1053}; \citealt{Levshakov:2005:827}), however the next large sample with strong statistical significance was not published until \citet{Webb:2011:191101} and \citet{King:2012:3370}.  This study was conducted on the Very Large Telescope's (VLT's) UVES instrument and analyzed 154 QSO absorption systems. \citet{Webb:2011:191101} found a statistically non-zero $\Delta\alpha/\alpha$ in the VLT sample, just as with the Keck sample, but its weighted mean value was greater, rather than smaller, than the laboratory value. This was most pronounced at high redshifts where $\Delta \alpha/\alpha$ took the same magnitude but with opposite sign to the \citet{Murphy:2004:131} Keck results.  Further, when \citet{Webb:2011:191101} and \citet{King:2012:3370} combined the Keck and VLT data-sets, evidence emerged for a variation in $\alpha$ across the sky. Because the Keck targets are primarily northern and the VLT targets primarily southern, together they cover most of the sky, enabling spatial variation in $\alpha$ to be analyzed. Several models were fit to this spatial variation but the simplest mode, a dipole in $\alpha$, gave a $4.2$-$\sigma$ preference (over a monopole) with the pole at $17.4\pm0.9$ hr right ascension and $-58\pm9$ deg.\ declination and an amplitude of $(10.2\pm 1.2) \times 10^{-6}$.

\begin{figure}[h]\vspace{0.0em}
\epsscale{1.00}\plotone{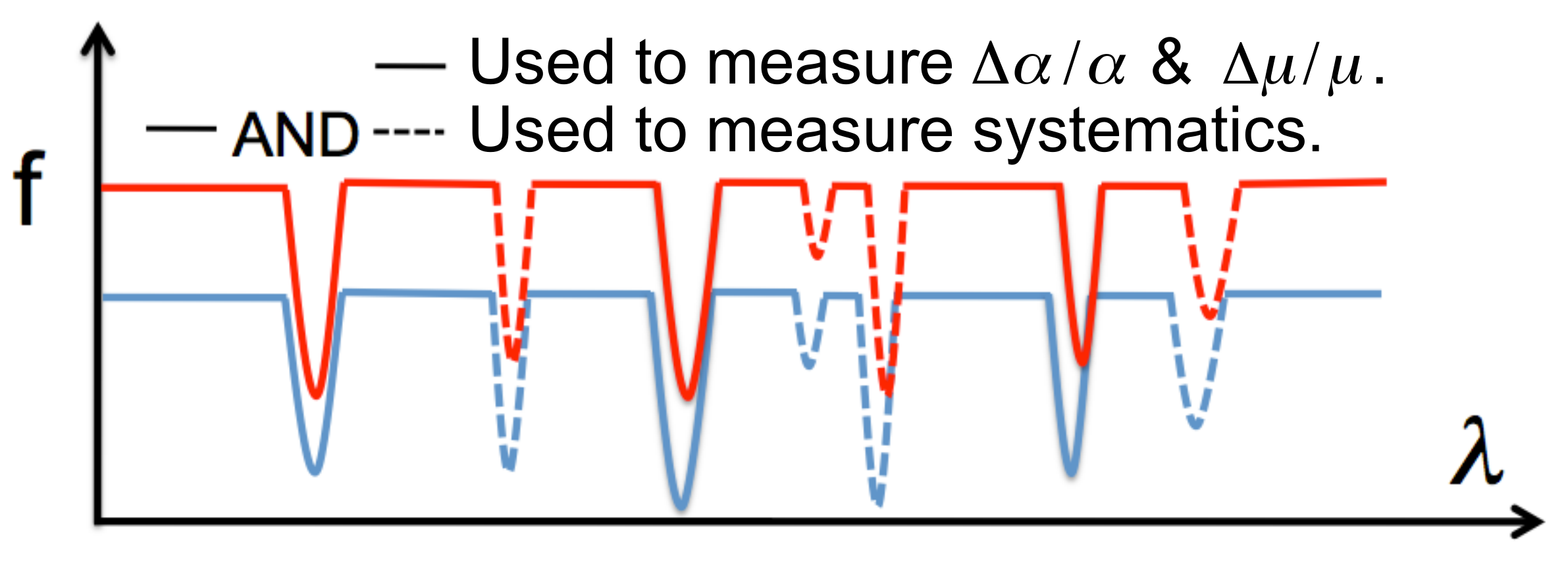}
\caption{\footnotesize Two spectra with wavelength scales distorted with respect to each other as a function of wavelength. Solid lines are examples of features that could be used to measure any $\alpha$-variation. However, \textit{all} spectral lines, dashed \textit{and} solid, can be used to measure the distortions between the spectra.}
\label{fig:mock_spec}
\end{figure}

Because the results from Keck give an average negative value for  $\Delta \alpha/\alpha$, while the results from the VLT yield a positive $\Delta \alpha/\alpha$, it is important to understand whether these measurements are an artifact of systematic errors between the telescopes.  In addition, if there are any \textit{velocity distortions} -- velocity shifts which change as a function of wavelength, see Figure~\ref{fig:mock_spec} -- present in a spectrum, as seen by, e.g., \citet{Griest:2010:158}, they could lead to a spurious measurement of $\Delta \alpha/\alpha$.  One of the most convincing methods to break the degeneracy between varying $\alpha$ (or $\mu$) and systematic errors is to observe equatorial targets on both telescopes and then compare the resulting \textit{spectra} (not simply $\alpha$ or $\mu$ values) from the two telescopes.  Some effort has already been made to make these comparisons, such as in \citet{King:2012:3370} and \citet{Wendt:2011:96}. 

\citet{King:2012:3370} measured velocity shifts between pairs of spectra from Keck and VLT using a Voigt profile fitting approach. Even though only some spectral lines are useful for measuring $\Delta\alpha/\alpha$, any well-defined feature contains information about measuring systematic shifts (Figure~\ref{fig:mock_spec}). \citet{King:2012:3370} fitted many of these `$\alpha$' and `non-$\alpha$ transitions' with Voigt profiles in Keck and VLT spectra of the same objects.  The advantage of this approach is a reliable measurement of velocity shift between the two spectra and a well-defined uncertainty estimate.  However, this process is extremely time intensive and depends on the fitted models.  If the transitions are fit differently, it is possible that the velocity shift derived would also be different.  Despite these limitations, \citet{King:2012:3370} applied this method to provide a first estimate of possible shifts between 7 pairs of VLT/UVES and Keck/HIRES spectra. In six pairs, the Keck and VLT data agreed well, however the seventh showed significant velocity distortions.

The other method of detecting shifts and distortions involves calculating the cross correlation function of regions in both spectra that contain features (e.g.~\citealt{Kanekar:2011:L12}, \citealt{Wendt:2011:96}).  The advantage of this method over the line-fitting approach is that it is completely model independent and non labor-intensive.  However, there are several short-comings of this method: many of the distortions that have been found (\citealt{Griest:2010:158} and \citealt{Whitmore:2010:89}) are on a sub-pixel level. The smallest shifts a simple cross-correlation approach can calculate are limited to the size of a pixel unless interpolation of the cross-correlation function is performed. Secondly, it requires additional time and effort to estimate an uncertainty in the velocity shift found.  For example, to estimate the uncertainty on velocity shifts in this way, \citet{Kanekar:2011:L12} cross-correlated 10000 pairs of simulated spectra with the structure and noise properties of the actual spectra.  Applying this approach to many features, possibly in many pairs of spectra, would become difficult and time-consuming. It also introduces a model dependency into the error estimate.

In this paper, we introduce a new method to directly compare spectra of the same object, hereafter referred to as the Direct Comparison (DC) method.  As further explained in section~\ref{sec:DC_method}, this method has the objectivity and model-independence of the cross-correlation technique, as well as the robustness and reliable uncertainty estimate of the line-fitting method used by \citet{King:2012:3370}.  In the past, the only way to reduce sensitivity to systematic effects in individual spectra was to analyze large samples of absorbers, like those in \citet{Murphy:2004:131} and \citet{King:2012:3370}.  In a large enough sample, many systematics average out \citep{Whitmore:2010:89}, but in a small set of spectra, systematic velocity distortions could more easily produce spurious detections of varying fundamental constants. Our new means of directly comparing two spectra should help establish reliability in small samples of absorption systems, by allowing discovery of, and possibly removal of systematic velocity distortions, whatever their origin.

In the rest of the paper we present the details of the DC method followed by its application to the Keck and VLT spectra of a QSO that have previously been used to measure $\Delta \mu/\mu$.  In section~\ref{sec:DC_method} we detail the method, including: how it avoids the problems present in the line-fitting and cross-correlation approaches, the DC method's basic formalism, and remaining limitations.  In section~\ref{sec:DC_J2123} we apply this method to the quasar J2123$-$0050.  We then apply the DC method to look for any systematics between the two telescopes and present our findings. A quasar used to measure $\Delta \mu/\mu$ is an ideal target to demonstrate the DC method because there are so many features present in the Lyman-$\alpha$ forest and there is also likely a large number of metal lines present, redwards of the forest, associated with the molecular hydrogen absorption system. Finally, we make some concluding remarks in section~\ref{sec:DC_conclusions} and discuss the other possible applications of the DC method such as well as radial velocity searches for extra-solar planets and real-time observation of the expansion of the Universe.

\section{The Direct Comparison Method}
\label{sec:DC_method}

\subsection{Basic formalism}
The direct comparison (DC) method directly compares two spectra of the same object at the same wavelengths using a $\chi^{2}$ minimization to find the best-fit velocity shift between them. It can be used to compare different exposures from the same telescope or spectra of the same object observed with different telescopes.  The ultimate goal of this method is to find any velocity distortions or velocity shifts between spectra and remove these velocity distortions over the whole spectrum.  This method requires only three user-defined parameters to be set: the scale of a simple Gaussian smoothing applied to the spectra, the size of the spectral regions to be directly compared (and for which an individual velocity shift measurement will be made), and the significance threshold above which  regions containing reliable velocity shift information will be selected.  The distribution of these individual velocity shifts can be used to decide the best method to correct for velocity shifts and distortions across the whole spectrum. Since there is no a priori knowledge of what the functional form of the distortions will be, the functional form of the final correction will depend on the results measured by the DC method. In addition, the DC method produces a reliable velocity shift and a reliable uncertainty estimate.  It also has the advantage that one does not first need to identify features in the spectra, or which features are sharp and narrow enough to provide the best results.

Consider two spectra to be directly compared, $f_1(i)$ and $f_2(j)$, which are dispersed onto different arrays of pixels, $i$ and $j$, and do not necessarily have the same velocity dispersion or resolution.  To compare the two spectra and allow for a sub-pixel velocity shift between them, we used the {\sc barak} package \footnote{Written and maintained by N.~Crighton at \url{https://pypi.python.org/pypi/Barak}.} to convolve the spectra with a Gaussian of constant full width at half maximum (FWHM) in velocity space. Smoothing with a velocity kernel roughly equal to the instrument resolution allows us to increase the signal-to-noise ratio (SNR) as high as possible without significantly degrading the resolution of the spectra or losing spectral information.  The size of this smoothing kernel is the first of the tunable parameters for the DC method. A natural choice for the FWHM is roughly equal to the resolution element of the telescopes, i.e.~2--3 pixels.  

After smoothing, we interpolate the flux and error array of one spectrum with a cubic spline (we use `spline' colloquially as a verb).  Therefore, we convert the flux, $f_2(j)$, and error, $\sigma_{2}(j)$, from a discrete spectrum to continuous functions of velocity, $v$, and user-specified parameters $\vec{p}$ -- $f_m(v,\vec{p})$ and $\sigma_{m}(v,\vec{p})$ -- where the extra parameters $\vec{p}$ are the parameters of the $\chi^2$ minimization procedure discussed below, including the main parameter of interest here, the velocity shift between the spectra.  Ultimately it is better to spline the spectrum with more `information' in it so as to maximize the accuracy of the spline.  For pairs of spectra with similar resolution, it is preferable to spline the higher SNR spectrum but, if one spectrum has a smaller velocity dispersion, it may be better to spline it instead.

The spectra are broken into velocity `chunks' -- sub-sections of the spectra of user-specified velocity length -- and the chunk size is the the second tunable parameter in the DC method. Chunks may overlap with each other to any user-specified extent. One might consider overlapping the chunks if looking for distortions over short wavelength scales at the expense of having independent measurements.  However, in this work we only consider independent, non-overlapping chunks. We discuss the velocity length in the context of real examples in sections \ref{sec:results_red} and \ref{sec:results_blue}.  Each of the chunks in one spectrum is compared to the corresponding chunk in the other.  We then minimize a modified $\chi^2$ statistic to determine the best-fit velocity shift between corresponding chunks:
\begin{equation}
\chi^2= \sum\limits_{i}\frac{\left[f_{1}(i)-f_{m}(i)\right]^2}{\sigma_{1}^2+\sigma_{m}^2}\,.
\label{eq:chi2}
\end{equation}
In our implementation we minimize $\chi^2$ via a Levenberg--Marquardt method where the free parameters are a velocity shift, a relative flux scaling, and a tilt between the pair of chunks. In this equation, $f_1$ refers to the flux of the non-splined spectrum, $f_m$ is the flux of the continuous, splined spectrum, while $\sigma_1$ and $\sigma_m$ are their corresponding error arrays. After $\chi^2$ is minimized we get a value and an uncertainty for the velocity shift, amplitude, and tilt, calculated from the relevant diagonal terms of the covariance matrix.  The parameter in which we are most interested is the velocity shift, however the other free parameters allow for possible differences in how the data were normalized and reduced.

The decision about how to set the velocity length of the `chunks' is based on several considerations.  One determining factor is to minimize any differences in the shape of the continuum over the chunk length so that only first order corrections (scaling and tilt) are necessary to accurately compare the chunks. Smaller chunks help in this regard.  However, the minimum chunk size should be larger than the extent of typical absorption features so that the velocity shift and slope parameters fitted in the DC method are not degenerate. Finally, in trying to decide a maximum size for velocity chunks, we aim to have as many independent chunks as possible. That is, while in principle the user may choose a very large chunk size, that may limit the understanding gained about any short-range or even some longer-range wavelength distortions present in the spectra. A secondary consideration is that, at lower SNRs, a larger chunk contains more potential for the noise fluctuations in the flux to influence the velocity shift estimate and its uncertainty -- see section \ref{sec:remaining_lims} for an exploration of low SNR extremes. Another major benefit to breaking the spectra into chunks is that it allows the DC method to be applied to the whole spectrum without first identifying regions of absorption features: once the process is complete, regions dominated by well-defined features have small error bars relative to those with fewer or broader features (Figure$~$\ref{fig:depth_test}).

\subsection{Monte Carlo testing}

\begin{figure*}[]\vspace{0.0em}
\epsscale{1.0}\plottwo{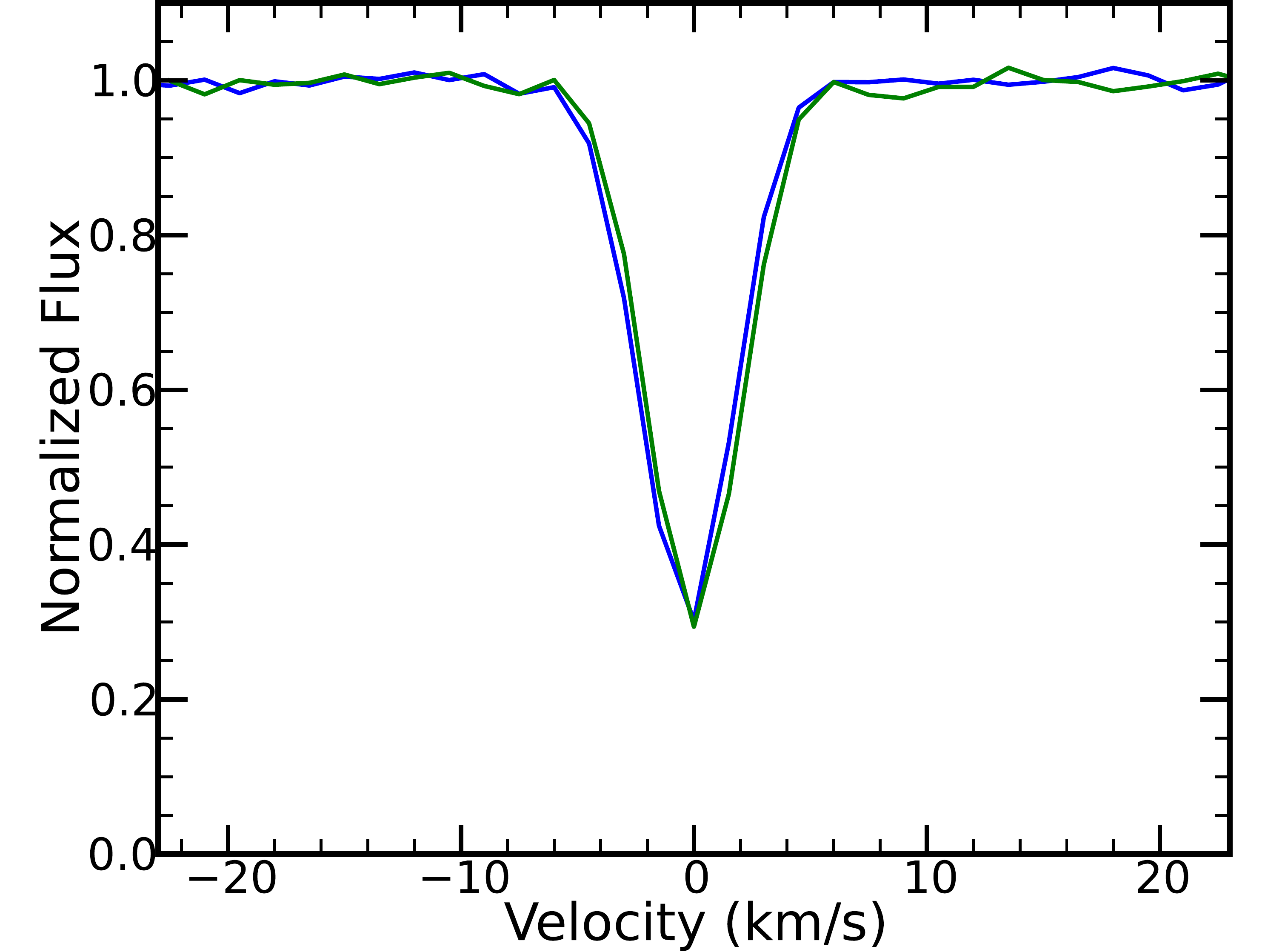}{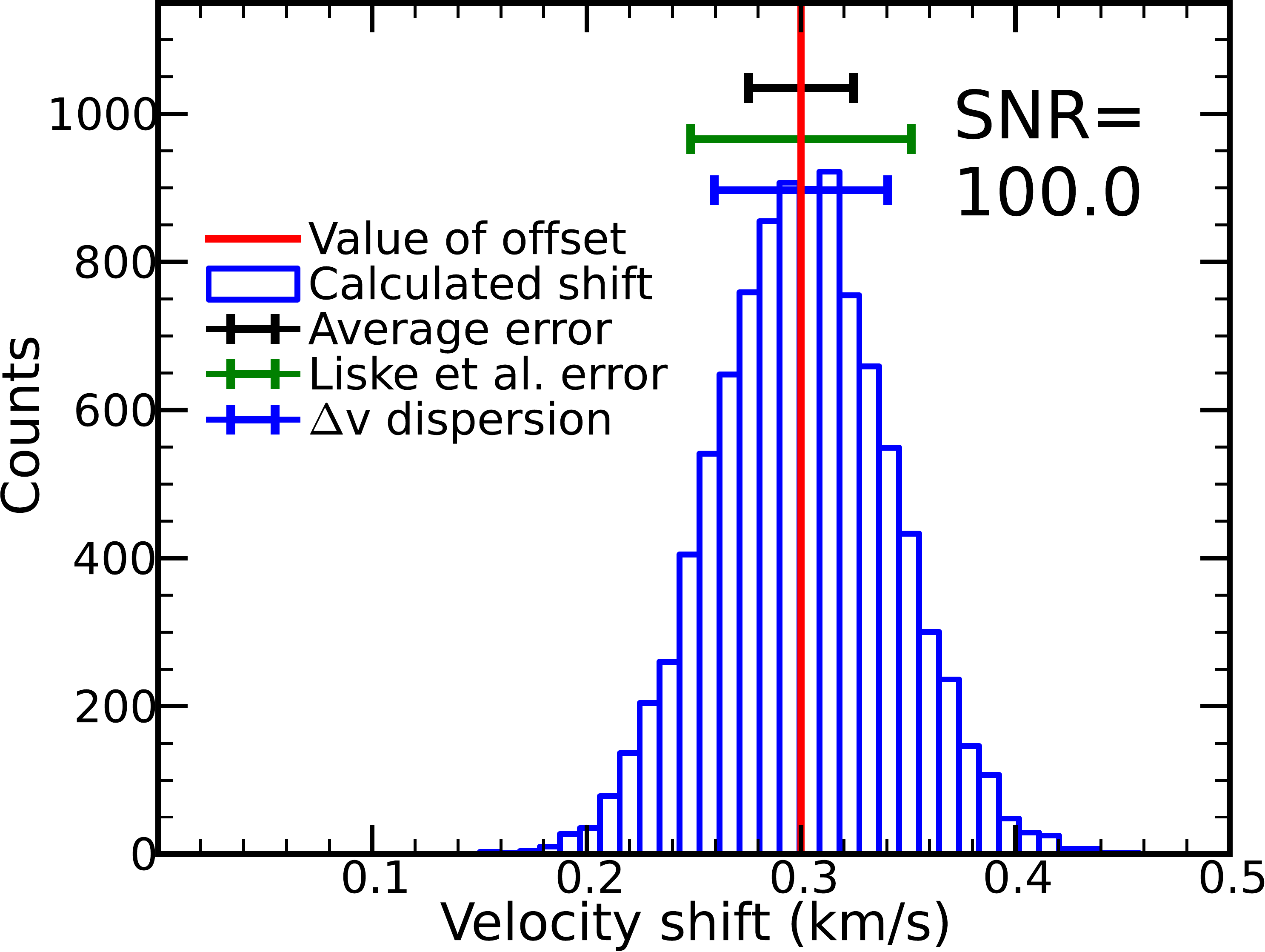}
\caption{\footnotesize Basic Monte Carlo test of the DC method with high SNR spectral features. The left panel shows simulated spectra with a SNR of 100 and an offset of 0.3\,km\,s$^{-1}$. The Gaussian absorption features have a depth of 0.7 and 1-$\sigma$ width of 2\,km\,s$^{-1}$. The dispersion of both spectra is 1.5\,km\,s$^{-1}$\,pix$^{-1}$. The right panel shows the distribution of velocity shifts found by the DC method in 5000 realizations of the left panel's spectra with different realizations of noise.  The vertical red line represents the known, correct shift. The black bar represents the mean uncertainty on the shifts from all realizations as calculated by the $\chi^2$ minimization. The green bar is the average Liske et al.~uncertainty. The blue bar represents the 1-$\sigma$ width of the distribution. Notice that the uncertainty measurements are not representative of the 1-$\sigma$ width of the Gaussian; this stems from the smoothing process and is addressed further in section~\ref{sec:smooth_cor} }
\label{fig:SNR100_hist}
\end{figure*}

To test the reliability of the DC method we produced synthetic spectra and ran them in pairs through a series of Monte Carlo simulations.  Consider two such spectra with the same absorption line profile but shifted with respect to each other.  We represent these spectra as single velocity chunks of unit continuum with a single Gaussian absorption feature. Each simulation involves two spectra with the same feature but with different random Gaussian noise.  We then apply a sub-pixel shift between the two spectra followed by the DC method to see how well we can recover that shift.  Figure~\ref{fig:SNR100_hist} shows an example of the simulated spectra and the resulting distribution of velocity shifts from a Monte Carlo test with 5000 realizations. These spectra have a SNR of 100 per pixel in the continuum, a velocity dispersion of 1.5\,km\,s$^{-1}$ pix$^{-1}$, are Gaussians with a $\sigma$ of 2.0\,km\,s$^{-1}$ and are offset by 0.3\,km\,s$^{-1}$. The reason that we choose a $\sigma$ for the Gaussian that is so small is that many metal lines in high resolution spectra are unresolved.  Therefore, it is natural to choose a width of simulated spectra to be similar to what we would see as an unresolved metal line. The right-hand panel of Figure~\ref{fig:SNR100_hist} shows that there is a roughly Gaussian distribution of recovered velocity shifts around the correct shift of 0.3\,km\,s$^{-1}$. However, we find unreliable uncertainties on the measured velocity shifts.  This problem occurs because of correlations between the fluxes of neighboring pixels, but it is simple to address and we detail the small corrections required in section \ref{sec:smooth_cor}.  Even with this problem with the uncertainty, this simple test demonstrates that the DC method recovers (i) the correct sub-pixel velocity shift and (ii) a Gaussian distribution of velocity shift measurements, at least at relatively high SNR.

\subsection{Corrections due to smoothing}
\label{sec:smooth_cor}
The Gaussian smoothing of the spectra causes the uncertainty in individual velocity shift measurements (derived from the $\chi^2$ minimization) to be smaller than the Monte Carlo distribution of velocity shifts in Figure~\ref{fig:SNR100_hist} (right panel). There are two factors contributing to this problem.

\begin{figure}[]\vspace{0.0em}
\epsscale{1.00}\plotone{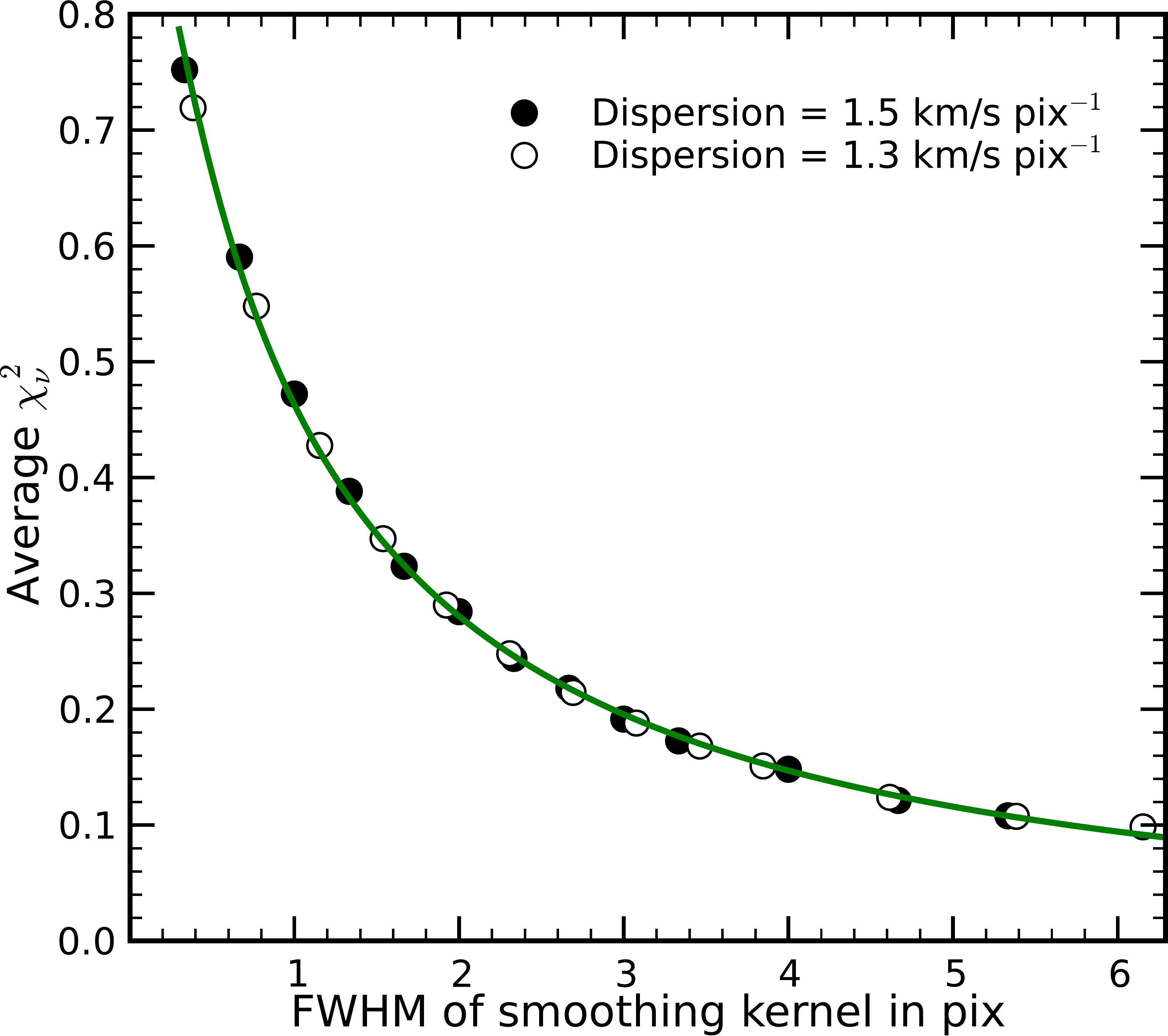}
\caption{\footnotesize The best fit to $\chi^2$ as a function of the number of pixels in the smoothing kernel. We find the best fitting function to be an exponential quadratic (equation \ref{eq:xi}) of the number of pixels used in smoothing the spectrum.  By converting to pixels instead of km\,s$^{-1}$, we are able to find a single equation that represents all spectra regardless of their dispersion.}
\label{fig:chi_smoothing}
\end{figure}

First, by smoothing the flux arrays of the spectra, the statistical noise per pixel is reduced, and this must be reflected in a correction to the error arrays. Without that correction, the smoothing leads directly to a reduced $\chi^2$, $\chi^2_{\nu}$, from the DC method which is too small. By repeating the simulations represented in Figure~\ref{fig:SNR100_hist} with different SNR and FWHM of the smoothing kernel, we find that, to first order, $\chi^2_{\nu}$ does not depend on the SNR and only depends on the FWHM.  As such, we can model $\chi^2_{\nu}$ as function of smoothing size, as shown in Figure~\ref{fig:chi_smoothing}), to derive a correction factor, $\xi_\chi$, for the error arrays of the spectra. From Figure~\ref{fig:chi_smoothing} we find best approximation for the underestimation of $\chi^2_{\nu}$ to be
\begin{equation}
	\xi_\chi = \exp(-0.16x^2-0.62x-0.76),
	\label{eq:xi} 
\end{equation}
where $x$ is the FWHM of the smoothing kernel in number of pixels.  After finding the correction value, $\xi_{\chi}$, we then multiply the corresponding error array by the square root of $\xi_\chi$ prior to performing the $\chi^2$ minimization in the DC method.  This modifies equation~\ref{eq:chi2} to:
\begin{equation}
\chi^2= \sum\limits_{i}\frac{\left[f_{1}(i)-f_{m}(i)\right]^2}{\xi_{1\chi}\left(\sigma_{1}\right)^2+\xi_{m\chi}\left(\sigma_{m}\right)^2}\,,
\label{eq:chi_cor}
\end{equation}
where $\xi_{1\chi}$ and $\xi_{m\chi}$ are the corresponding correction factors to the un-splined error array and the splined error array, respectively.  With these corrections made, we obtain a reduced $\chi^2$ value of approximately unity when analyzing smoothed, simulated spectra.

\begin{figure}[]\vspace{0.0em}
\epsscale{1.00}\plotone{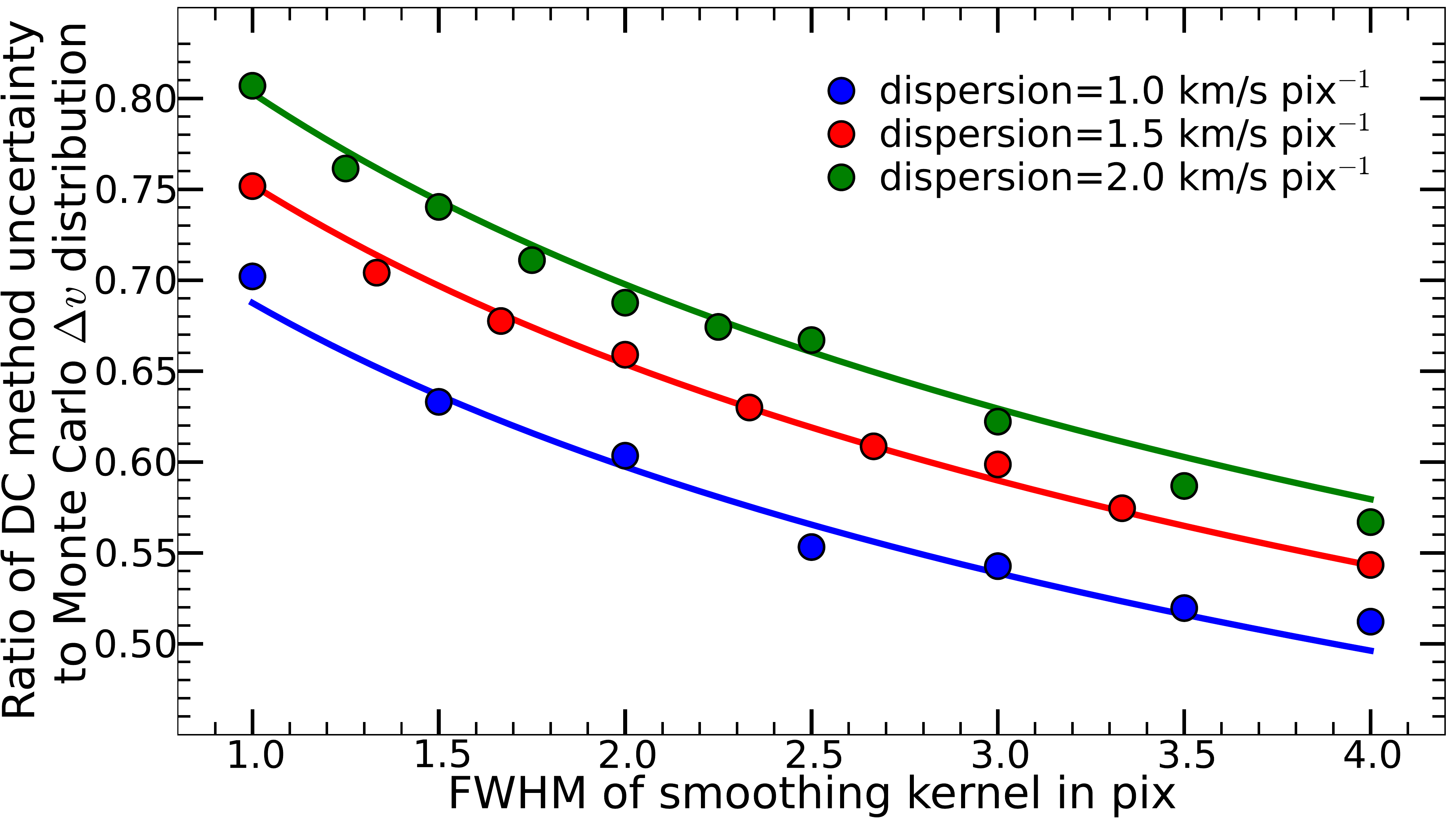}
\caption{\footnotesize The correction to our DC method uncertainties for simulated spectra with three different dispersions but constant resolution.  The correction curves shown here are calculated by multiplying equations~\ref{eq:xi_sig} and~\ref{eq:xi_FWHM}. The small scatter around the curve implies that the DC method correctly determines the error on the velocity shift to within a few percent accuracy.}
\label{fig:uncert_cor}
\end{figure}

The second correction we must make is to the uncertainty measured in the velocity shift between chunks, $\sigma_{\Delta v}$, derived from the $\chi^2$ minimization.  The correction in the previous step ($\xi_\chi$) only compensates for the over-estimation of the error arrays because of the smoothing applied to the flux arrays. However, smoothing the fluxes also introduces extra covariance between pixels that is not accounted for by the previous correction. These correlations between pixels cause the root-mean-square (RMS) flux variations to be slightly smaller than implied by the error array, even after correction in the previous step. To determine a correction for this, we model the ratio of the uncertainty calculated from the $\chi^2$ minimization to the 1-$\sigma$ width of the Monte Carlo velocity shift distribution, $\sigma_{\Delta v}/\Delta v_{\rm MC}$, for a given SNR, as a function of $x$, the FWHM (in pixels) of the smoothing kernel. Unlike the previous correction, we find that this ratio depends on both $x$ and the instrumental resolution, ${\rm FWHM}/2\sqrt{2 ln(2)}$, in units of the spectral dispersion, which we define as $y$, i.e.~$y\equiv\left({\rm FWHM}/2\sqrt{2 ln(2)}\right)/{\rm dispersion}$. We find that the dependence of the ratio $\sigma_{\Delta v}/\Delta v_{\rm MC}$ on $x$ and $y$ is separable; that is, it can be expressed as $\xi_\sigma \times \xi_{\rm FWHM}$ where $\xi_\sigma$ is function of $x$ only and $\xi_{\rm FWHM}$ is a function of $y$ only. We find that the best-fitting expressions for $\xi_\sigma$ and $\xi_{\rm FWHM}$ are:
\begin{equation}
	\xi_{\sigma} = \exp(-0.05x^2-0.17x-0.28), 
	\label{eq:xi_sig}
\end{equation}
and
\begin{equation}
	\xi_{\rm FWHM} = -0.15y+1.22.
	\label{eq:xi_FWHM}
\end{equation}
Figure~\ref{fig:uncert_cor} shows values of $\sigma_{\Delta v}/\Delta v_{\rm MC}$ for a range of $x$ which is appropriate for applications to real spectra while holding the resolution of the simulated spectra constant. Over-plotted is the total correction, $\xi_{\sigma} \times \xi_{\rm FWHM}$, as a function of $x$, for the three different dispersions. The three different curves represent the three values of $y$. For the typical values of $y$ explored here ($0.8\lesssim y \lesssim 2.0$) we recover a DC method uncertainty that is correct to within a few percent; this is apparent in the relatively small scatter around the three curves in Figure~\ref{fig:uncert_cor}.

It is worth pointing out that the corrections from equations~\ref{eq:xi} and~\ref{eq:xi_FWHM} (seen in Figures~\ref{fig:chi_smoothing} and~\ref{fig:uncert_cor}) do not appear to approach exactly unity as the smoothing kernal size approaches 0\,km\,s$^{-1}$.  If there were no smoothing at all, we would expect there to be no need for the correction functions, though we do not force our fits to obey this boundary condition. However, we have chosen only to model the corrections needed over a range of practical smoothing values.  Smoothing kernels smaller than 1 pixel are not recommended as there is no additional information present on a sub-pixel level.  Likewise, smoothing kernels larger than a few pixels are also not recommended because there will be a greater loss of spectral information.

With this correction, we can convert the uncertainty measured from the $\chi^2$ minimization, $\sigma_{\Delta v}$, into a more accurate and statistically meaningful uncertainty on the velocity shift between two spectral chunks:
\begin{equation}
  \sigma^\prime_{\Delta v} = \frac{\sigma_{\Delta v} \times \sqrt{\chi^2_{\nu}}}{\xi_{\sigma} \times \xi_{\rm FWHM}}. 
\label{eq:sigma_cor}
\end{equation}
If there were no correlations between pixels (such as in an un-smoothed simulation) we would expect $\sigma^\prime_{\Delta v}$ to equal $\sigma_{\Delta v}$. However, since there is always some correlation between pixels in real spectra, even before we smooth the flux arrays, we must also multiply our final uncertainty estimate by $\sqrt{\chi^2_{\nu}}$. Thus, equation~\ref{eq:sigma_cor} provides the final velocity uncertainty estimate from the DC method.

\begin{figure}[]\vspace{0.0em}
\epsscale{1.00}\plotone{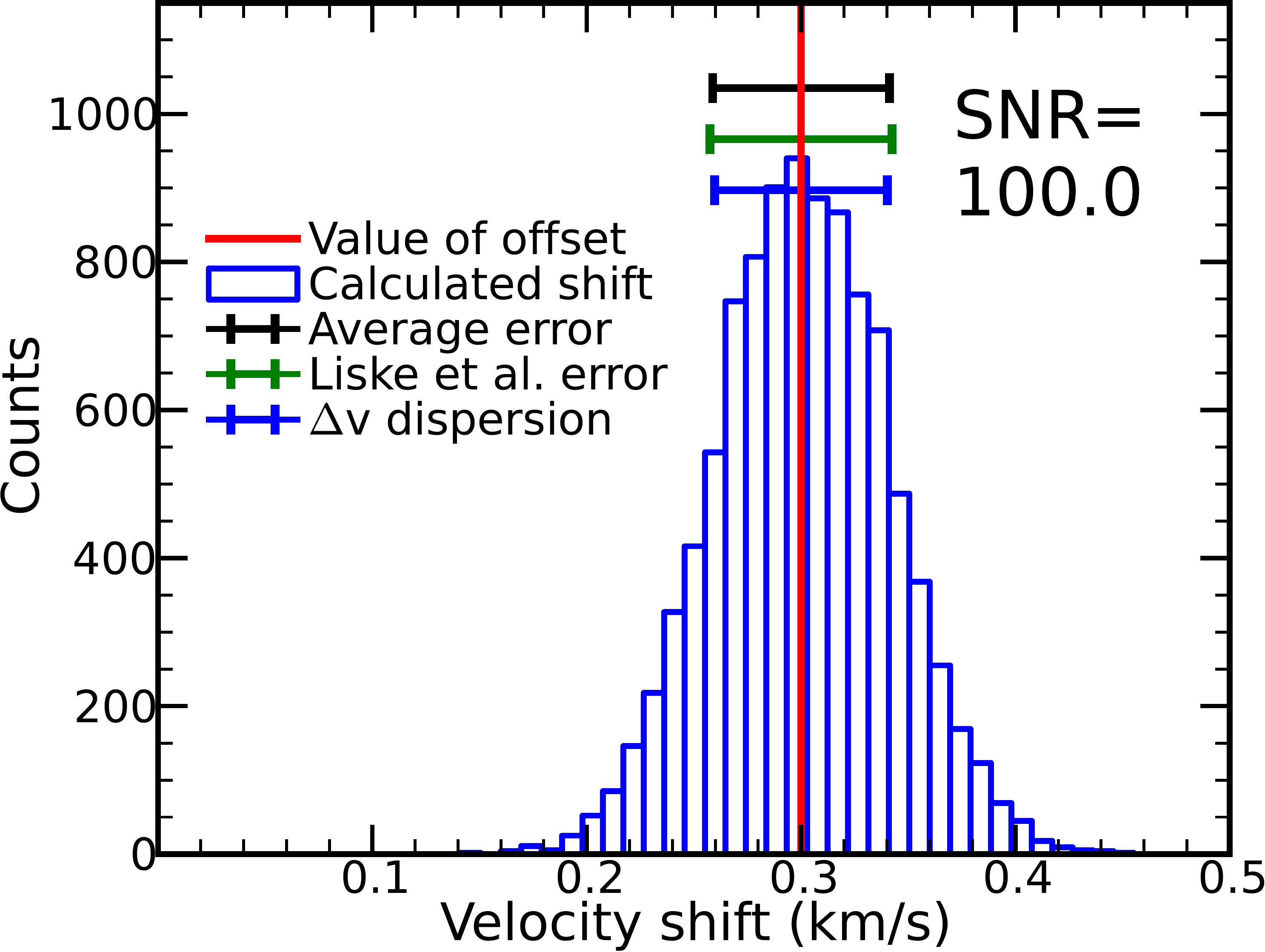}
\caption{\footnotesize Results from a Monte Carlo test with the same input parameters as those used for Figure~\ref{fig:SNR100_hist} but corrected for the effects of smoothing using equations~\ref{eq:chi_cor} \& \ref{eq:sigma_cor}. The lines and horizontal bars are color-coded as in Figure~\ref{fig:SNR100_hist} (right panel). After applying the correction to the error arrays of the spectra (equation~\ref{eq:chi_cor}) and to the DC method uncertainty estimate (equation~\ref{eq:sigma_cor}), we find that it matches well the width of velocity shift distribution.}
\label{fig:hist_cor}
\end{figure}

Figure~\ref{fig:hist_cor} shows another Monte Carlo simulation with all of the same input parameters as those used to create Figure~\ref{fig:SNR100_hist}, with the exception that the corrections from equations~\ref{eq:chi_cor} and~\ref{eq:sigma_cor} have been applied.  This gives a reduced $\chi^2$ of $\chi^2_\nu \approx 1$ as well as an uncertainty roughly equal to the distribution of the measured velocity shifts.

\subsection{Objective selection of reliable features}
\label{sec:feat_select}

The third, and final, parameter that must be tuned in the DC method is a means of selecting which chunks are dominated by noise and which contain reliable measurements. If the SNR of the spectra, which clearly plays a major role in determining the uncertainty on the $\Delta v$ measurements, were constant across the entire wavelength range of interest, we could select such regions simply by choosing a maximum velocity shift uncertainty and rejecting all $\Delta v$ measurements with uncertainties larger than this threshold. However, in reality, the SNR of echelle spectra changes significantly over the optical wavelength range, showing a particularly strong fall-off at progressively bluer wavelengths than $\sim$4000\,\AA. A chunk must contain enough sharp features to return a velocity shift uncertainty somewhat smaller than the ``saturation value'' for that chunk size, the point where the noise fluctuations are contributing a comparable amount to $\chi^2$ as the features. In our implementation of the DC method, we calculate a ``tracker point'' using simulations for each chunk to objectively determine whether that chunk contains enough spectral information -- enough sharp features with a high enough SNR -- to provide a reliable velocity shift measurement with a reliable uncertainty estimate. We describe this technique below.

We first determine how much information is in each chunk via a method outlined by \citet{Liske:2008:1192}. They provide a means of calculating the uncertainty on $\Delta v$ between two (chunks of) spectra, determined by how much spectral information is present in the chunk.  This velocity uncertainty, $\sigma_{v_i}$,  is calculated by the following equation for each pixel, $i$:
\begin{equation}\label{eq:Liske}
	\sigma^2_{v_{ i}} = \left( \frac{1}{\frac{dS_i}{dv}} \right)^2 \left[ \sigma_{1i}^2 + \sigma_{2i}^2 + \frac{\left( S_{2i} - S_{1i} \right)^2 }{ {\left( \frac{c}{v}\right)^2} \left( \frac{dS_i}{dv}\right)^2} \sigma^2_{S'_i}  \right]\,. 
\end{equation}
This is dependent on the uncertainty in flux, the gradient in flux and in the uncertainty of the gradient of the flux at each pixel in the pair of spectra. $S$ is the flux, $dS_i/dv$ is the average gradient in the flux at the $i^{th}$ pixel, and $\sigma^2_{S'_i}$ is the average uncertainty on $dS_i/dv$ between the two spectra. The Liske et al.~uncertainty on $\Delta v$ for an entire chunk is then the sum over all of pixels in that chunk:
 \begin{equation}\label{eq:liskesum}
	\sigma^2_{v} =  \frac{1}{\sum\limits_{i} \sigma^{-2}_{v_{ i}}}\,.  
\end{equation}
Calculating this Liske et al.~uncertainty requires that both spectra be on the same pixel grid, so it is therefore always calculated after smoothing the spectra and the spline interpolation has been performed. We express $dS_i/dv$ as an averaged, one-sided finite-difference derivative:
\begin{equation}
  dS_i/dv = \frac{\left( \frac{\left(S_{i+1}-S_{i} \right)_1}{\sigma^2_1} + \frac{\left(S_{i+1}-S_{i} \right)_2}{\sigma^2_2} \right)}
                      {\left(\frac{1}{\sigma_1^2} + \frac{1}{\sigma_2^2}\right )}/dv.
\end{equation}
Propagating the errors in the above equation gives an expression for $\sigma^2_{S^\prime_i}$,
 \begin{equation}
    \sigma^2_{S'_i} = \left(\frac{1}{\sigma_{1i}^2} + \frac{1}{\sigma_{2i}^2} \right)^{-2} \left( \frac{\sigma_{1i+1}^2 + \sigma_{1i}^2}{\sigma_{1i}^2} + \frac{\sigma_{2i+1}^2 + \sigma_{2i}^2}{\sigma_{2i}^2} \right),
  \end{equation}
and the following expression for the Liske et al.~uncertainty on any pixel:
  \begin{equation}
 \sigma^2_{v_{i}} = \left( \frac{v_{i+1}-v_{i}}{dS} \right)^2 \left( \sigma_{1i}^2 + \sigma_{2i}^2 + z\right),
  \end{equation}
  where
   \begin{equation}
   z=\frac{\left( S_{2i} - S_{1i} \right)^2 }{\left( \frac{S_{1{i+1}} - S_{1{i}}}{\sigma^2_{1i}} + \frac{S_{2{i+1}} - S_{2{i}}}{\sigma^2_{2i}} \right)^2}  \left( \frac{\sigma_{1i+1}^2 }{\sigma_{1i}^2} + \frac{\sigma_{2i+1}^2}{\sigma_{2i}^2 } + 2 \right).
   \end{equation}
The sum over all the pixels in the chunk, as per equation~\ref{eq:liskesum}, then yields the \citeauthor{Liske:2008:1192}~error for the chunk.

\begin{figure}[]\vspace{0.0em}
\epsscale{1.0}\plotone{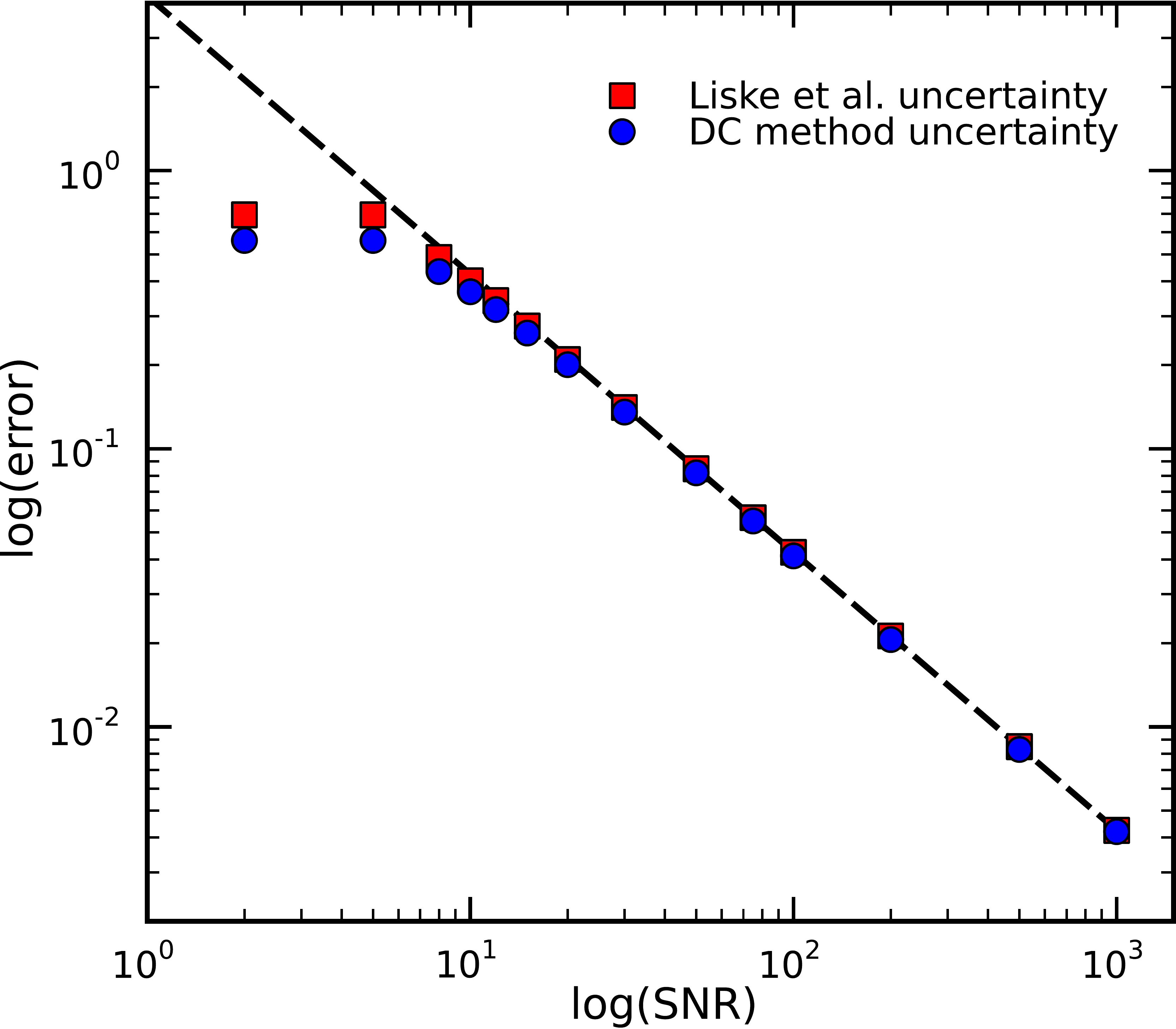}
\caption{\footnotesize Reliability of the DC method as a function of signal-to-noise ratio (SNR).  Simulated spectra were generated as in Figure~\ref{fig:SNR100_hist} (left panel) with a range of SNR values in the continuum.  The mean 1-$\sigma$ uncertainty on the velocity shift returned by the DC method (over 10000 realizations) for each SNR value is plotted as a blue circle.  The mean \citeauthor{Liske:2008:1192}~uncertainty (see equation~\ref{eq:Liske}) for each SNR value is plotted as a red square.  The solid lines extrapolate those mean uncertainties at the highest SNR by assuming that $\sigma(\Delta v) \propto$ SNR$^{-1}$.  Note that the DC method underestimates the true uncertainty at SNR $\lesssim7$ per pixel.}
\label{fig:err_vs_SNR}
\end{figure}

Our aim is to use the above Liske et al.~formalism, applied to synthetic realizations of a chunk, to determine whether that chunk contains enough spectral information to yield a reliable $\Delta v$ measurement, i.e.~whether or not that chunk's spectral features, if any, are sharp and/or numerous enough to dominate contributions to $\chi^2$ over those from noise fluctuations alone. From Figure~\ref{fig:err_vs_SNR} it is clear that the Liske et al.~velocity shift uncertainty shows the same behavior as that determined using the DC method: it `saturates' at $\rm SNR\la7$\,pix$^{-1}$, indicating that, for a feature of relative depth 0.7 (see Figure~\ref{fig:SNR100_hist} left-hand panel), the noise, not the feature, dominates $\chi^2$ at lower SNR. We simulate what an average Liske et al.~uncertainty would be for a given pair of chunks with no feature present and compare this value to the Liske et al.~uncertainty calculated from the real pair of chunks.  This allows us to select any chunks with Liske et al.~uncertainty significantly lower than measurements from corresponding simulated chunks of continuum. These selected points will have reliable measurements of velocity shifts and accurate uncertainties.

In practice, our approach is as follows. For each chunk of the real spectra we create 50 synthetic realizations of that chunk. The flux in each of these synthetic spectra is generated as random Gaussian noise with an average value of unity.  The standard deviation of the Gaussian noise, as well as the error array, is set to a constant value which is the median value of the error array from the corresponding chunk in the real spectra. The mean Liske et al.~uncertainty and root-mean-square (RMS) deviation then provide a ``tracker point'' for that chunk: if the Liske et al.~uncertainty in the real chunks is $\le N$ times the RMS below the tracker point, we accept that chunk as having sufficient spectral information to provide a reliable $\Delta v$ measurement. The number, $N$, times the RMS of the tracker points is the final parameter set by the user.  Therefore, choosing a higher $N$ will be more selective in determining which features to use, while a lower $N$-value will accept chunks with less information.

\begin{figure*}[t]\vspace{0.0em}
\epsscale{1.00}\plotone{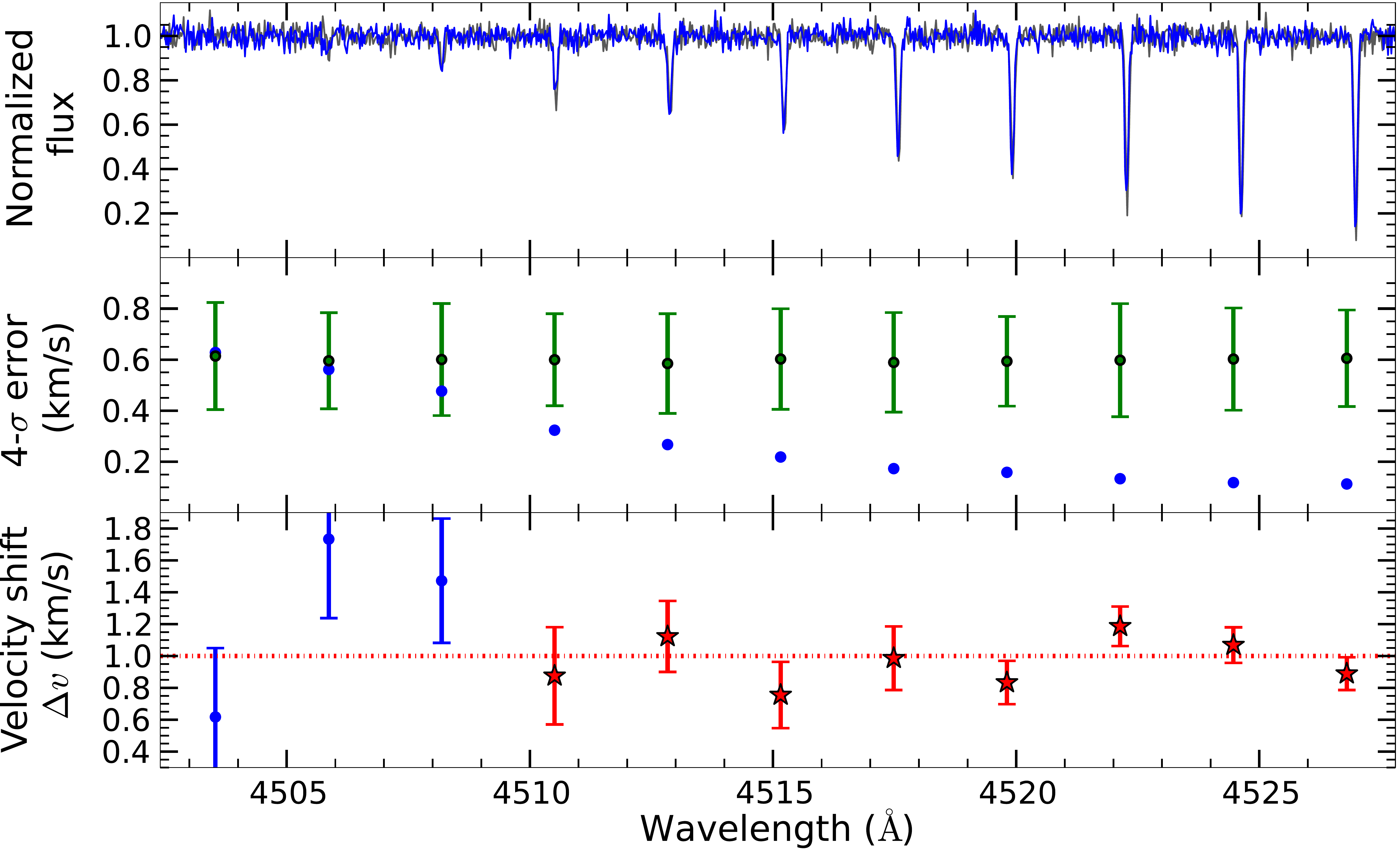}
\caption{\footnotesize Features of increasing depth and their corresponding tracker points.  The top panel shows the two simulated spectra with decreasing feature depths.  The middle panel shows the Liske et al.~uncertainty of the chunks in blue and the tracker points with $N=4$ in green.  The bottom panel shows the measured velocity shifts for each chunk and the applied offset of 1.0 \,km\,s$^{-1}$ as a red dashed line. Here we can see that since the error array is the same for all chunks, the tracker points are level across the spectrum.  As the depth of the feature in the chunk increases, the Liske et al.~uncertainty for the chunk drops below the 4$\sigma$ cutoff imposed and starts to produce a velocity shift measurement consistent with the applied velocity shift and with a accurate uncertainty measurement.}
\label{fig:depth_test}
\end{figure*}

We test this method with simulations that can be seen in Figure~\ref{fig:depth_test}.  We simulate spectra in the same fashion as before but this time apply a 1.0 \,km\,s$^{-1}$ velocity shift between the two spectra and allow for chunks of different depths. The middle panel of Figure~\ref{fig:depth_test} shows the ``tracker point'' calculated via the simulations for each chunk compared with the Liske et al.~uncertainty from the real spectra. As the feature depth increases, the spectral information increases and the Liske et al. uncertainty decreases. Once it decreases below $N=4$ times below the Liske et al.~uncertainty range from the simulations, the lower panel demonstrates that the velocity shift measurement becomes more reliable (and precise).

\subsection{Generalization to more complex spectra}
\label{sec:complex}

In previous sections we address single, unresolved features in simulated spectra.  However, real spectra can contain much more complicated structure.  In this section we show that the uncertainty estimate on the velocity shift found by the DC method is robust to more complex absorption structure. From equation~\ref{eq:Liske}, in the case of a single feature per chunk, we can see that as the feature becomes broader the Liske et al.~uncertainty becomes larger, in linear proportion, because the gradient in flux decreases. Conversely, as the feature increases in depth (and the SNR in the continuum remains the same) the Liske et al.~uncertainty becomes smaller, again in linear proportion until the features saturates, after which no additional information is added.  Similarly, in the case of multiple, unblended absorption features, the uncertainty decreases with the square-root of the number of features.  These scalings stem from the fact that the Liske et al.~uncertainty accurately reflects the spectral information available. And, as demonstrated by \citet{Liske:2008:1192}, this is true even for much more complex spectra with many blended features of varying depth and width, e.g.~real Lyman-$\alpha$ forest spectra.

\begin{figure*}[t]\vspace{0.0em}
\epsscale{1.00}\plotone{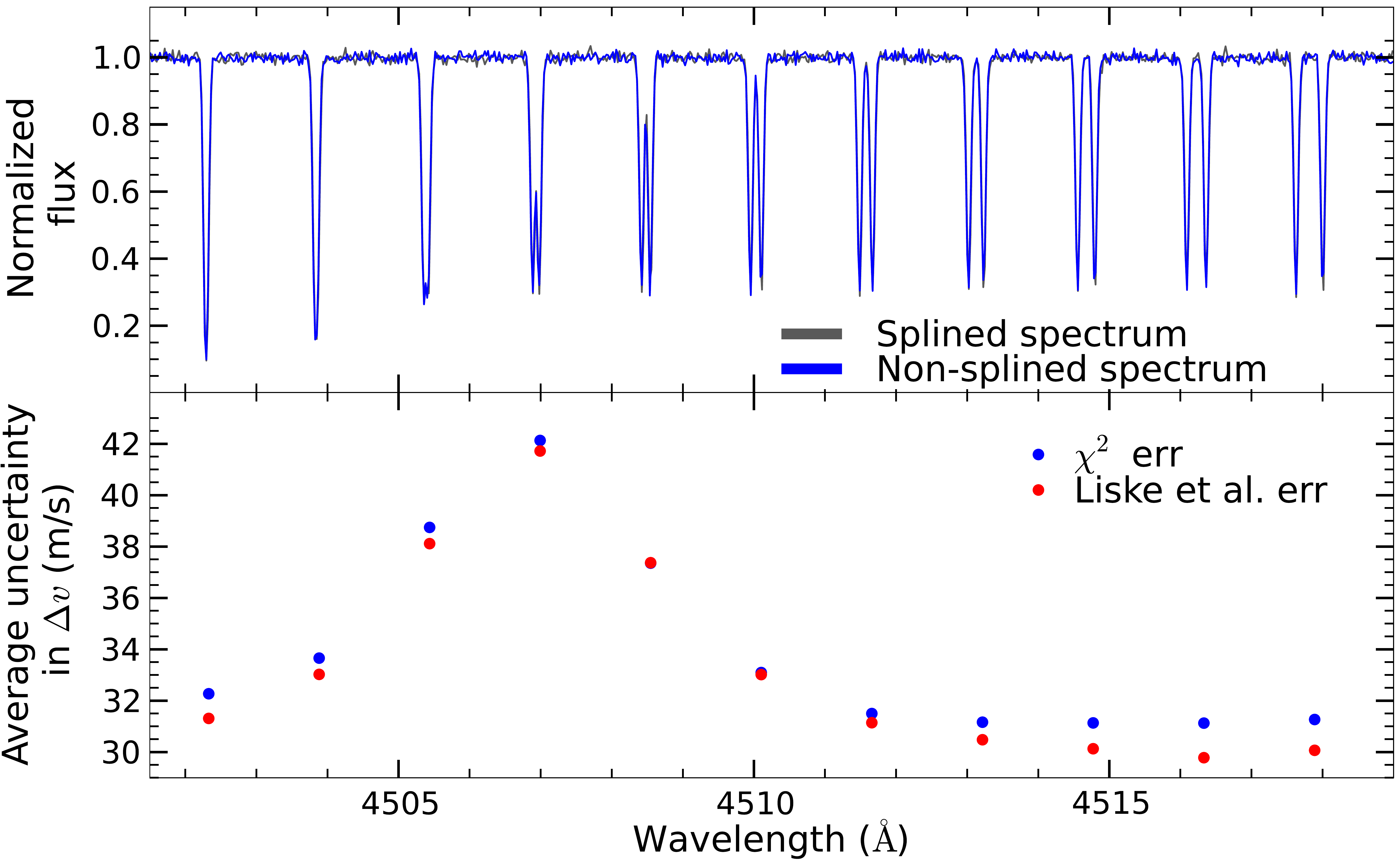}
\caption{\footnotesize The accuracy of the DC method uncertainty estimate with more complicated absorption profiles. The top panel shows simulated spectra with two features at different spacings from one another; each pair of features is centered in a chunk. The bottom panel presents the average $\chi^2$ minimization uncertainty (blue) and Liske et al.~uncertainty (red) for 5000 Monte Carlo simulations of the spectra from the top panel. That the DC method uncertainty tracks the Liske et al.~uncertainty closely in all cases demonstrates the accuracy of the former.}
\label{fig:twofeat}
\end{figure*}

Figure~\ref{fig:twofeat} demonstrates that the uncertainty measured from the $\chi^2$ minimization in the DC method traces the Liske et al.~uncertainty for varying feature width, depth and degree of blending of two spectral features. In the case of two independent spectral features (far right side of Figure~\ref{fig:twofeat}) we see that the uncertainties are relatively small, reflecting the increased information present, and very similar to the Liske et al.~uncertainties.  The chunk on the far left shows the two features entirely blended together, again resulting in small DC method and Liske et al.~uncertainties.  Because the features are on top of each other, the depth of the feature increases (lowering the uncertainty) but also broadens slightly (increasing the uncertainty).  We can see that, combined, this leads to an uncertainty slightly larger than two independent features; again, this closely reflects the behavior of the Liske et al.~uncertainties. In the central chunks of Figure~\ref{fig:twofeat} the two measures of uncertainty increase together to a maximum when the two features are blended such that they produce a broad feature without gaining much depth.

These simulations demonstrate that the DC method uncertainty is accurate because it traces well the Liske et al.~uncertainty when features broaden, deepen and blend. The former should therefore prove reliable once absorption structure is generalized even further to complex structure like the Lyman-$\alpha$ forest spectra upon which the Liske et al.~approach has previously been demonstrated. It also supports our use of the Liske et al.~uncertainty in simulations for selecting chunks with sufficient spectral information for reliable velocity shift measurements.

\subsection{Remaining limitations}
\label{sec:remaining_lims}

There are limitations to the DC method at low SNR.  To demonstrate this, 10000 realizations of pairs of chunks were created with SNR ranging from $3\le$ SNR $\le1000$. Figure~\ref{fig:err_vs_SNR} shows that the uncertainty on $\Delta v$ is underestimated with the DC method for a chunk with SNR $\lesssim7$ per pixel in the continuum.  Therefore, the DC method cannot deliver reliable information about the velocity shifts between two spectra at these low SNR values. However, the tracker points, described above are designed exactly for this reason -- to only select chunks with accurate uncertainties.  It is worth recalling that the other methods, like line-fitting and cross-correlation, will also fail at low SNR, so the DC method is not alone in this short-coming.

\begin{figure*}[]\vspace{0.0em}
\epsscale{1.0}\plottwo{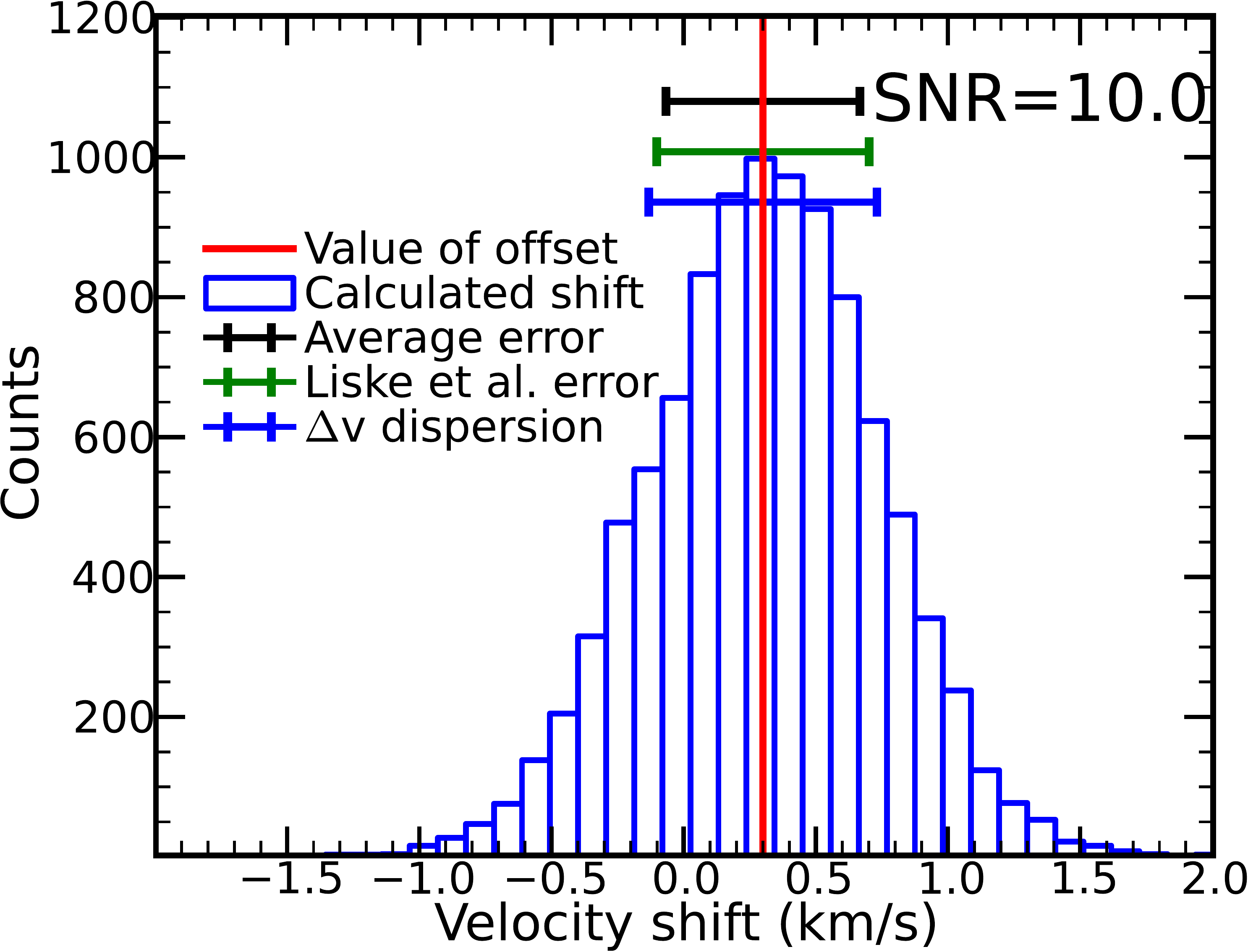}{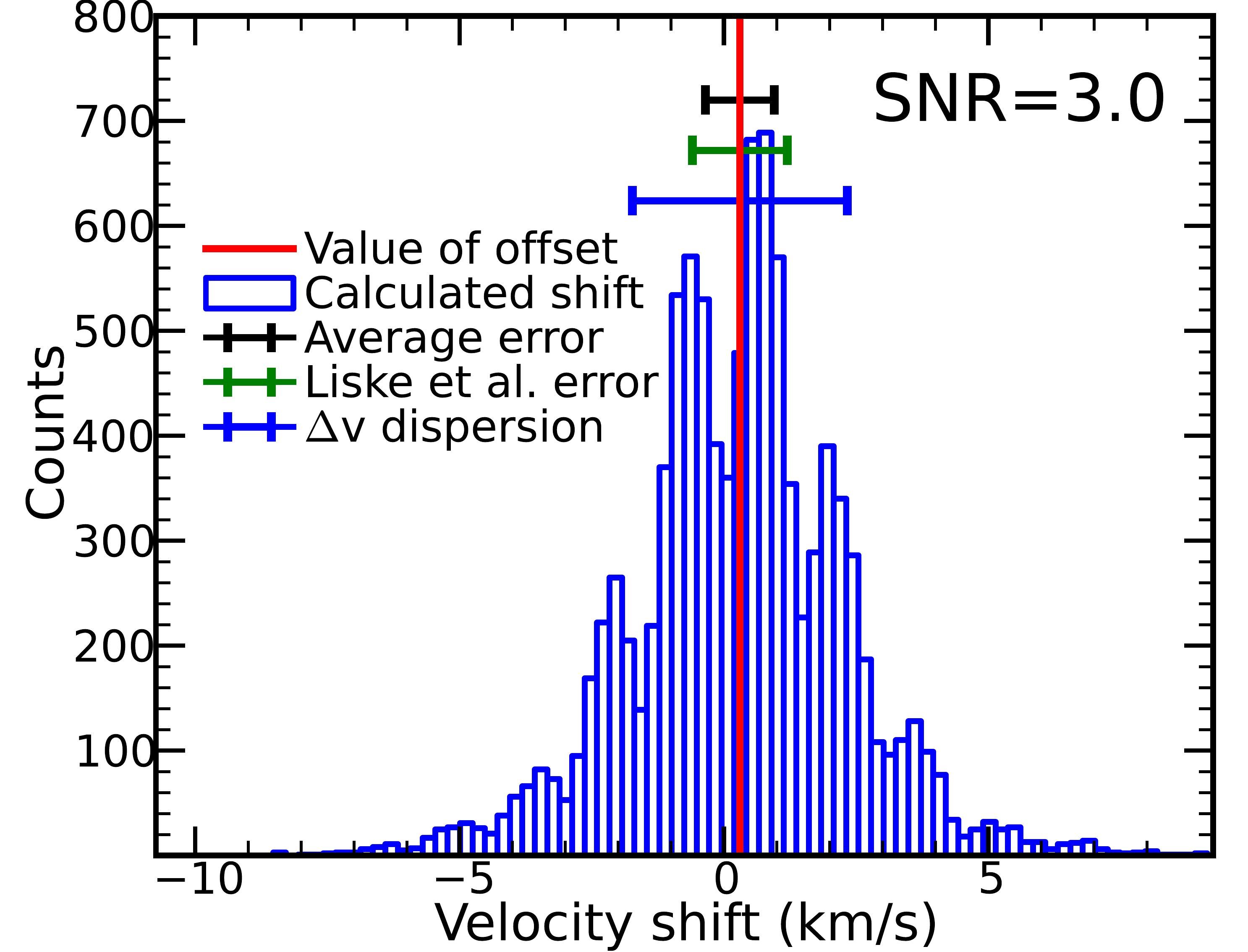}
\caption{\footnotesize Examples of the breakdown of the DC method at low SNR. In both plots, the DC method is applied to 10000 realizations of the absorption feature in Figure~\ref{fig:SNR100_hist} (left panel). The left panel has a SNR$\sim 10$ per pixel in the continuum and the right panel has a SNR$\sim 3$. The lines and horizontal bars are color-coded as in Figure~\ref{fig:SNR100_hist} (right panel).  In the left case, we show that at lower SNRs the uncertainty measured from the $\chi^2$ minimization, as well as the Liske et al.~uncertainty, slightly underestimate the distribution of velocities found.  This is also reflected in Figure~\ref{fig:err_vs_SNR}. It is interesting to note, that the distribution is still Gaussian even if the uncertainties are underestimated.  In the right panel we see a more extreme failure of the DC method that can occur when the chunk is dominated by noise. Note that the distribution prefers a half-pixel shift and therefore provides neither a reliable measurement or reliable uncertainty when a chunk is dominated by noise.}
\label{fig:break_down}
\end{figure*}

While it is not recommended to trust any chunk with an underestimated uncertainty (i.e.~falling within $N$ times the simulation RMS of the tracker points), there is an even more extreme break-down of the DC method at lower SNRs.  The left-hand panel of Figure~\ref{fig:break_down} shows a Monte Carlo simulation of 10000 pairs of spectra at a SNR of 10 per pixel in the continuum.  While a chunk with SNR of 10 will likely be rejected, it is interesting to note that we still recover a roughly Gaussian distribution around the correct (0.3 \,km\,s$^{-1}$) velocity shift.  However, in the case of very low SNR (right-hand panel of Figure~\ref{fig:break_down}) we see that even the distribution is non-Gaussian.  Rather, there seems to be an artificial preference for half-pixel shifts.  This is most likely due to the need for interpolating the flux and error arrays in one spectrum; in our implementation of the DC method, we used a spline. As the splined error array is an overestimate of the actual uncertainty away from pixel centers, there will be a small dip in $\chi^2$ when half-pixel shifts are applied between the pair of (chunks of) spectra. Regardless of the cause, it is important to note that chunks with such low SNR will never be accepted by even the most lenient selection of an $N$ value for the tracker points.

Summarizing the results of these tests, the Direct Comparison method appears to be a robust tool that can measure sub-pixel velocity shifts between spectra of the same object.  It is model-independent, meaning that there is no bias from user input, and does not require any prior demarkation of spectral features.  Since it can be applied quickly and automatically across a pair of spectra in chunks, it can be used to find velocity shifts between each pair of chunks and also find changes in these shifts as a function of wavelength, i.e.~spectral distortions.  Moreover, the DC method provides a robust estimate of the uncertainty in its velocity shift measurements.  However, the DC method does have its limitations at SNR lower than $\sim 7$ per pixel in the continuum.  As long as this is kept in mind, the DC method can be applied to real spectra to better understand systematic errors between telescopes or epochs of observation.  The next section explores the DC method as applied to the QSO J2123$-$0050.

\section{Application to J2123\boldmath{$-$}0050}
\label{sec:DC_J2123}

SDSS J212329.46-005052.9, referred to in this paper as J2123$-$0050, is one of only six quasars to have yielded constraints on $\Delta \mu/\mu$.  To date, four quasars including J2123$-$0050 have provided a constraint on $\Delta \mu/\mu$ using H$_2$ and HD transitions, in absorbers at redshifts $z>2$, which fall within with the Lyman-$\alpha$ forest (e.g.~\citealt{King:2008:251304, Thompson:2009:1648, Wendt:2012:69}). Compared to most damped Lyman-$\alpha$ systems, those containing detectable column densities of molecules are likely much colder and denser and hence much more rarely detected, meaning few opportunities for measuring $\Delta\mu/\mu$.  Variations in $\mu$ should cause a shift in the rovibronic transitions of molecular hydrogen, with different transitions shifting by different amounts and in different directions; i.e.~the transitions have a variety of sensitivity coefficients, $K$, in equation 2.  The other two quasars have been observed in the radio and millimeter bands to reveal ammonia and, more recently, methanol at $z<1$. These molecules' transitions are 2 orders of magnitude more sensitive to variations in $\mu$ than H$_2$ and so have yielded very tight, null constraints ($\sim10^{-7}$) on $\Delta\mu/\mu$ (\citealt{Flambaum:2007:240801, Murphy:2008:1611, Kanekar:2011:L12, Jansen:2011:100801, Ellingsen:2012:L7, Bagdonaite:2013:46}).

\subsection{The Keck/HIRES and VLT/UVES spectra}
J2123$-$0050 is a bright ($r=16.45$\,mag) quasar at redshift $z_{\rm em}=2.261$ which was identified in the Sloan Digital Sky Survey data release 3 \citep{Abazajian:2005:1755}. The presence of so many spectral lines of varying degrees of depth and width make J2123$-$0050 an ideal target on which to test the reliability of the DC method.  The HIRES observations were performed by \citet{Milutinovic:2010:2071} in 2007 with a seeing of 0.3--0.5\,arcsec, resulting in a spectrum that covers 3071--5869\,\AA\ and has a SNR in the continuum of $\sim$25 per 1.3\,km\,s$^{-1}$ pixel at $\sim$3420\,\AA\ and roughly 40 at $\sim$5000\,\AA.  The UVES spectrum was observed by \citet{Weerdenburg:2011:180802} in 2008 with wavelength coverage of 3047--9466\,\AA\ and a SNR in the continuum of roughly 65  per 1.5\,km\,s$^{-1}$ pixel at $\sim$3420\,\AA\ and 60 at $\sim$5000\,\AA. There are 86 H$_2$ lines and seven HD lines over the wavelength range 3071--3421\,\AA. 

In the application of the DC method to J2123$-$0050, we only consider the final 1D spectra, i.e.~those resulting from the combination of many individual exposures. For each of the final HIRES and UVES spectra, the contributing exposures were combined without concern for overall velocity shifts or velocity distortions between them. The position of the quasar in the slit during a given exposure imparts an approximately constant velocity offset, and this offset will vary between exposures. These velocity offsets are only of secondary importance for varying fundamental constant measurements, as can be seen from equations \ref{eq:alpha} and \ref{eq:mu}, and generally only lead to a very slight broadening of narrow spectral features. However, velocity distortions between exposures would be a much more important systematic effect for such measurements. The DC method could be used to identify and quantify such velocity shifts and/or distortions between individual exposures from a single spectrograph. However, for the purposes of demonstrating the DC method in this paper, and for assessing the systematic effects on the previous measurements of $\Delta\mu/\mu$ from the HIRES and UVES spectra of J2123$-$0050, instead we only apply it to the final, combined 1D spectra here.

If there is a variation in $\mu$, the redder Lyman H$_2$ transitions should shift in the opposite direction to (and by similar amounts as) the bluer Lyman transitions, i.e.~the redder transitions' $K$-coefficients have opposite sign to those of the bluer transitions\footnote{A caveat is that the Werner lines at bluer wavelengths shift in the opposite direction to the bluest Lyman lines at similar wavelengths. In principle, this would mitigate the effect of any long-range velocity distortion on $\Delta\mu/\mu$. However,  \citet{Malec:2010:1541} and \citet{Weerdenburg:2011:180802} demonstrated that the Werner lines are sufficiently weak in these spectra that they do not influence $\Delta\mu/\mu$ significantly.}.  A long-range velocity distortion in one spectrum would, therefore, have a very similar effect on the relative velocities of the H$_2$ transitions as a varying $\mu$ and spuriously give a non-zero $\Delta \mu/\mu$. However, the transitions of atomic hydrogen in the Lyman-$\alpha$ forest should not vary with $\mu$, and give us a means to break the degeneracy between a variation in $\mu$ and long-range distortions. The presence of so many narrow, and likely unresolved ($\sim$6\,km\,s$^{-1}$-wide), H$_2$ transitions which constrain $\Delta\mu/\mu$, and the Lyman-$\alpha$ forest of relatively broad H{\sc \,i} lines which do not constrain $\Delta\mu/\mu$, make J2123$-$0050 a particularly interesting target on which to utilize the DC method.

While \citet{Malec:2010:1541} and \citet{Weerdenburg:2011:180802} used the spectra only for measuring $\Delta \mu/\mu$ from the H$_2$ and HD transitions in the Lyman-$\alpha$ forest, there are also well-defined metal transitions present redwards of the Lyman-$\alpha$ emission line that could, in principle, be used to measure $\Delta\alpha/\alpha$ in the damped Lyman-$\alpha$ system at $z_{\rm abs}=2.059$. There are also metal-line transitions at other redshifts, which would not usually be used to measure $\Delta\alpha/\alpha$.  The presence of both these types of lines allows the DC method to be applied redwards of the Lyman-$\alpha$ emission line. Thus, the Keck/HIRES and VLT/UVES spectra of J2123$-$0050 allow a comprehensive test for velocity distortions using the DC method over the wavelength range $\sim$3000--5200\,\AA.  Finally, one unusual feature of this QSO is that it has time-variable C{\sc \,iv} absorption systems falling just bluewards of the C{\sc \,iv} emission line in the wavelength range 4840--4925\,\AA\, \citep{Hamann:2011:1957}.  Therefore, the region containing these time-variable absorption systems must be masked out when calculating possible velocity distortions between the spectra.

\subsection{Results}
\subsubsection{Redwards of Lyman-$\alpha$ emission}
\label{sec:results_red}
\begin{figure*}[ht]\vspace{0.0em}
\epsscale{1.0}\plotone{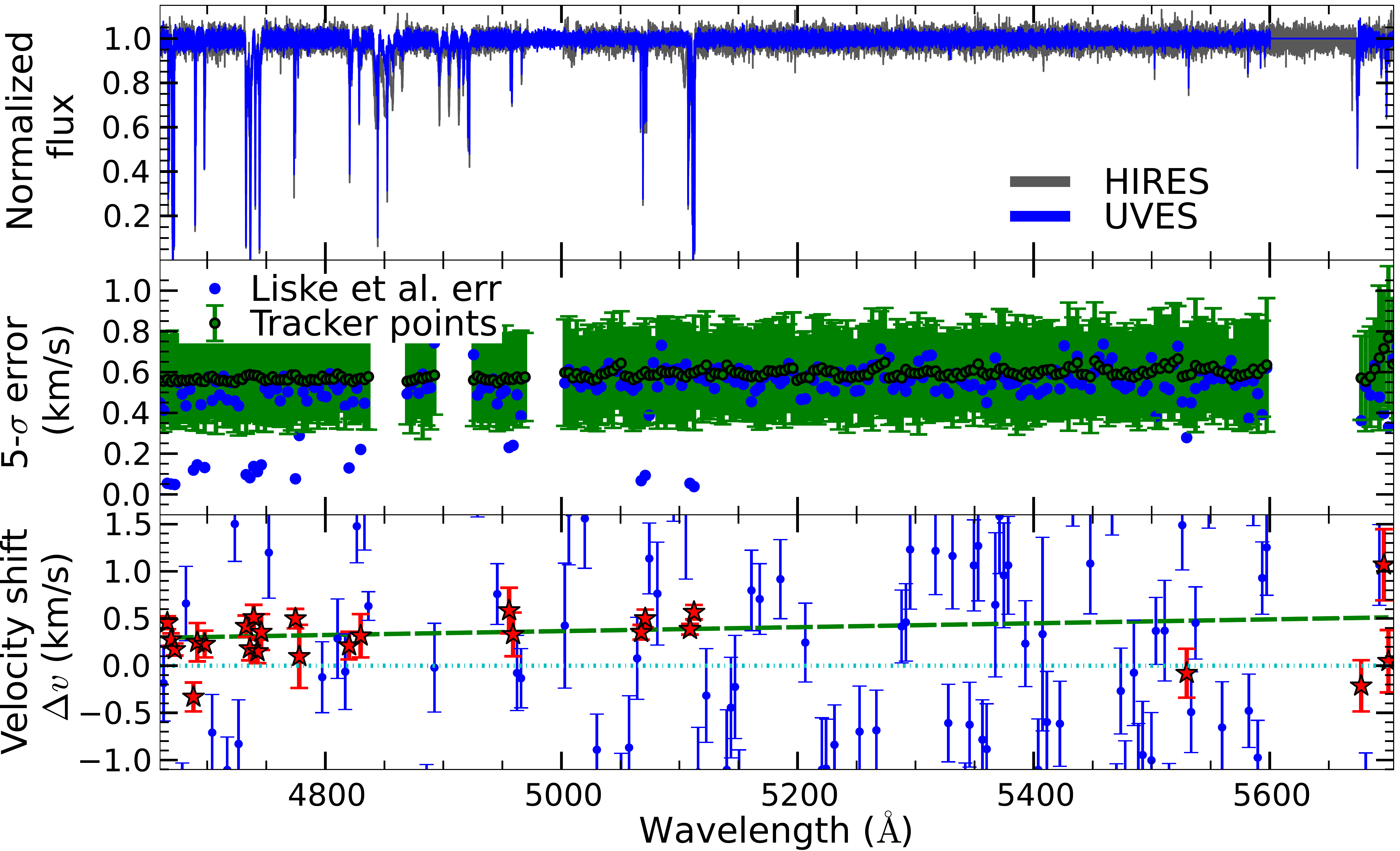}
\caption{\footnotesize Results from the DC method redwards of the Lyman-$\alpha$ emission line for J2123$-$0050. The top panel shows the spectra from both UVES (blue line) and HIRES (grey line).  The middle panel shows the \citet{Liske:2008:1192} uncertainty (equation \ref{eq:Liske}) for each 200\,km\,s$^{-1}$-wide chunk and their corresponding tracker points.  The error bars on the tracker points represent 5 times the RMS of the simulations. Chunks in which the Liske et al.~uncertainty is smaller than 5 times RMS of the simulations below the tracker points are likely to provide reliable velocity shift estimates and uncertainties from the DC method; they are `selected' for further analysis. Note that the chunks containing the time-variable C{\sc \,iv} absorption lines have been masked out.  The bottom panel shows the velocity shifts and their 1-$\sigma$ uncertainties derived from the DC method for all (unmasked) chunks.  The selected chunks are shown in red stars. Note the much smaller scatter in the velocity shift estimates for the selected points compared to the others.  The dashed green line is an weighted least squares fit to the selected chunks. }\label{fig:red_results}
\end{figure*}

Figure~\ref{fig:red_results} presents our results for the red portion of J2123$-$0050. The top panel shows the Keck and VLT spectra in the wavelength region containing a series of absorption features, $\sim$4650--5700\,\AA.  Note that while there are more features in J2123$-$0050 both to lower and higher wavelengths than those shown in Figure~\ref{fig:red_results}, only this wavelength region contains overlapping spectra from both telescopes.  The HIRES spectrum has a resolving power of $R\approx110000$ while the UVES spectrum has $R\approx60000$ in the red, meaning that the sharper metal lines are substantially less resolved in the UVES spectrum. This can be seen clearly in Figure~\ref{fig:red_results} where some features appear slightly shallower in the UVES spectrum than in the HIRES spectrum. Ideally, the DC method would be applied to pairs of spectra with the same resolution, so as not to lose much information in the convolution process. However, since our aim is to measure velocity shifts between the spectra (cf.~measuring metal column densities or Doppler broadening parameters), the information lost from smoothing should have only minor effects on the results. For the purposes of our demonstration of the DC method on J2123$-$0050, we do not attempt to quantify this here. Also, this region contains the aforementioned C{\sc \,iv} variable absorption features as well as a UVES chip gap, both of which have been masked in the analysis. At 5000\,\AA\, the SNR for the HIRES spectrum is $\sim$40 per 1.3\,km\,s$^{-1}$ pixel and for UVES it is $\sim$60 per 1.5\,km\,s$^{-1}$ pixel. These SNR values, coupled with the sharpness and multicomponent nature of the absorption features in this wavelength range, yield an average uncertainty in velocity shifts calculated by the DC method of $\sim$0.18\,km\,s$^{-1}$.  In this portion of the spectrum we find that most features are well defined and, at most, $\sim$150\,km\,s$^{-1}$. Because of this we choose a chunk size that is slightly larger, 200\,km\,s$^{-1}$, allowing us to cover whole regions of absorption but also maximize the number of measurements we can make in a limited wavelength range.

The middle panel of Figure~\ref{fig:red_results} shows how the reliable velocity shift measurements -- i.e.~those associated with features containing sufficient spectral information -- were discerned from those unlikely to be reliable.  \citeauthor{Liske:2008:1192}~uncertainties (blue points) for the real spectra that fall below 5 times the RMS of the tracker points (green points) correspond to reliable chunks to use for $\Delta v$ measurements.  It is reassuring to note that chunks that are selected via this method do correspond to regions with strong  features in both spectra.

The bottom panel of Figure~\ref{fig:red_results} summarizes the results from the red portion of the J2123$-$0050 spectra.  All velocity shifts and their 1-$\sigma$ uncertainties determined from the DC method are plotted, while the measurements selected as reliable are represented as red stars.  It is immediately clear that the points identified as reliable show a much smaller scatter in velocity than those associated with featureless continuum regions of the spectra. The reliable points show a general consistency in their velocity shift values, with a possible trend towards larger velocity shifts between the spectra at longer wavelengths. A weighted linear regression was fit to these selected points, giving an average offset of $0.31 \pm 0.05$\,km\,s$^{-1}$ and a slope of $(2.1 \pm 1.0)$\,m\,s$^{-1}$nm$^{-1}$ with a $\chi^2$ per degree of freedom ($\chi^2_{\nu}$) of 2.6. Because $\chi^2_{\nu} > 1$, a linear fit may not be the best model for the measured velocity shifts. Therefore, to obtain a more representative error estimate for the slope, we added a constant value in quadrature to the uncertainties of all the points such that $\chi^2_{\nu}$ reduced to unity. This effectively marginalizes over the extra scatter around the best fitting line.  Doing this gives us a best fitting line with a slope of $(0.4 \pm 1.7)$\,m\,s$^{-1}$nm$^{-1}$, corresponding to a null detection of distortions between the spectra.  However, this may not be the full story.  If the 4 features bluewards of $\sim 5000$ \AA\, are excluded from the fit then we measure a slope of $(4.9 \pm 2.2)$\,m\,s$^{-1}$nm$^{-1}$ over a range of 450 \AA, which is marginally non-zero. It is therefore possible that a velocity distortion does exist between the Keck and VLT spectra in this case, though it may have a complex variation with wavelength.  Finally, the overall offset of 0.31\,km\,s$^{-1}$ is likely due to the quasar, on average, being in a slightly different part of the slit when observed with Keck/HIRES and VLT/UVES and does not directly impact measurements of varying constants (see equations~\ref{eq:alpha} and~\ref{eq:mu}). The positive sign of this average offset means that the HIRES spectrum is redshifted relative to the UVES spectrum, on average.

Although we find no significant velocity distortion in our very simple (linear) fit above, it is interesting to understand how precisely the distortion measurement could constrain systematic errors in a varying constants analysis. As an example of a typical $\Delta \alpha/\alpha$ analysis, we imagine an absorber with metal-line transitions spanning the $\sim$4650--5100\,\AA\ wavelength range used in measuring the most extreme slope between the HIRES and UVES spectra (shown in Figure~\ref{fig:red_results}), and consider Mg{\sc \,ii} $\lambda$2796 and the two Fe{\sc \,ii} transitions with rest frame wavelengths of 2586 and 2600\,\AA.  If we assume that Mg{\sc \,ii} $\lambda$2796 falls at the red edge of the range in Figure~\ref{fig:red_results}, implying an absorption redshift of $z_{\rm abs}=0.815$, then the Fe{\sc \,ii} $\lambda$2586 line would fall at the blue edge.  These transitions therefore form a representative combination for $\Delta \alpha/\alpha$ studies from the wavelength range in Figure~\ref{fig:red_results}. These three transitions are commonly seen in quasar spectra and, because Mg{\sc \,ii} $\lambda$2796 is insensitive to $\Delta \alpha/\alpha$ ($q=211$\,cm$^{-1}$) and the Fe{\sc \,ii} lines are strongly dependent on $\alpha$ (averaged to give $q=1410$\,cm$^{-1}$) this combination of lines normally contributes strongly to a constraint on  $\Delta \alpha/\alpha$.  The $(4.9 \pm 2.2)$ m\,s$^{-1}$nm$^{-1}$ slope corresponds to a total velocity distortion of $182\pm 81$ \,m\,s$^{-1}$ between the UVES and HIRES spectra over the 370\,\AA\ wavelength range from the Fe{\sc \,ii} $\lambda$2586 transition to the Mg{\sc \,ii} $\lambda$2796 line. Equation~\ref{eq:alpha} implies that a distortion of this magnitude and sign would lead to a measurement of $\Delta \alpha/\alpha$ being $10.0\pm4.5$\,ppm greater in HIRES than UVES.

One particularly interesting aspect of these results is the precision with which we can measure/limit the distortion and its effect on $\Delta\alpha/\alpha$. In this example, that precision  (i.e.~4.5\,ppm) is smaller than the typical statistical uncertainty on $\Delta\alpha/\alpha$ for Mg/Fe{\sc \,ii} absorbers in \citet{Murphy:2004:131} and \citet{King:2012:3370} with spectra of similar SNR to those studied here. That is, the DC method is generally capable of utilizing enough spectral information in a quasar spectrum to put interesting limits on, or identify and measure, systematic velocity distortions which are important for varying $\alpha$ measurements. With a larger sample of pairs of quasar spectra, it would therefore be possible to identify systematic errors that are likely shared by all the pairs and understand their effect on $\Delta\alpha/\alpha$ with higher precision.

Of course, the main caveat to this conclusion is that this Mg/Fe{\sc \,ii} absorber is just one example of transitions used to measure $\Delta\alpha/\alpha$. If instead we had imagined a higher-redshift absorber, quite different transitions, perhaps with quite a different set of $\Delta Q$ values, would have fallen in the wavelength range covered by Figure~\ref{fig:red_results}. The systematic effect on $\Delta\alpha/\alpha$ caused by the velocity distortion (assuming it is real; recall that it only has a 2-$\sigma$ statistical significance) would then be quite different, perhaps even of the opposite sign. Nevertheless, the above discussion highlights the invaluable information about systematic errors that it is possible to derive using the DC method.

\subsubsection{Bluewards of Lyman-$\alpha$ emission}
\label{sec:results_blue}

\begin{figure*}[]\vspace{0.0em}
\epsscale{1.0}\plotone{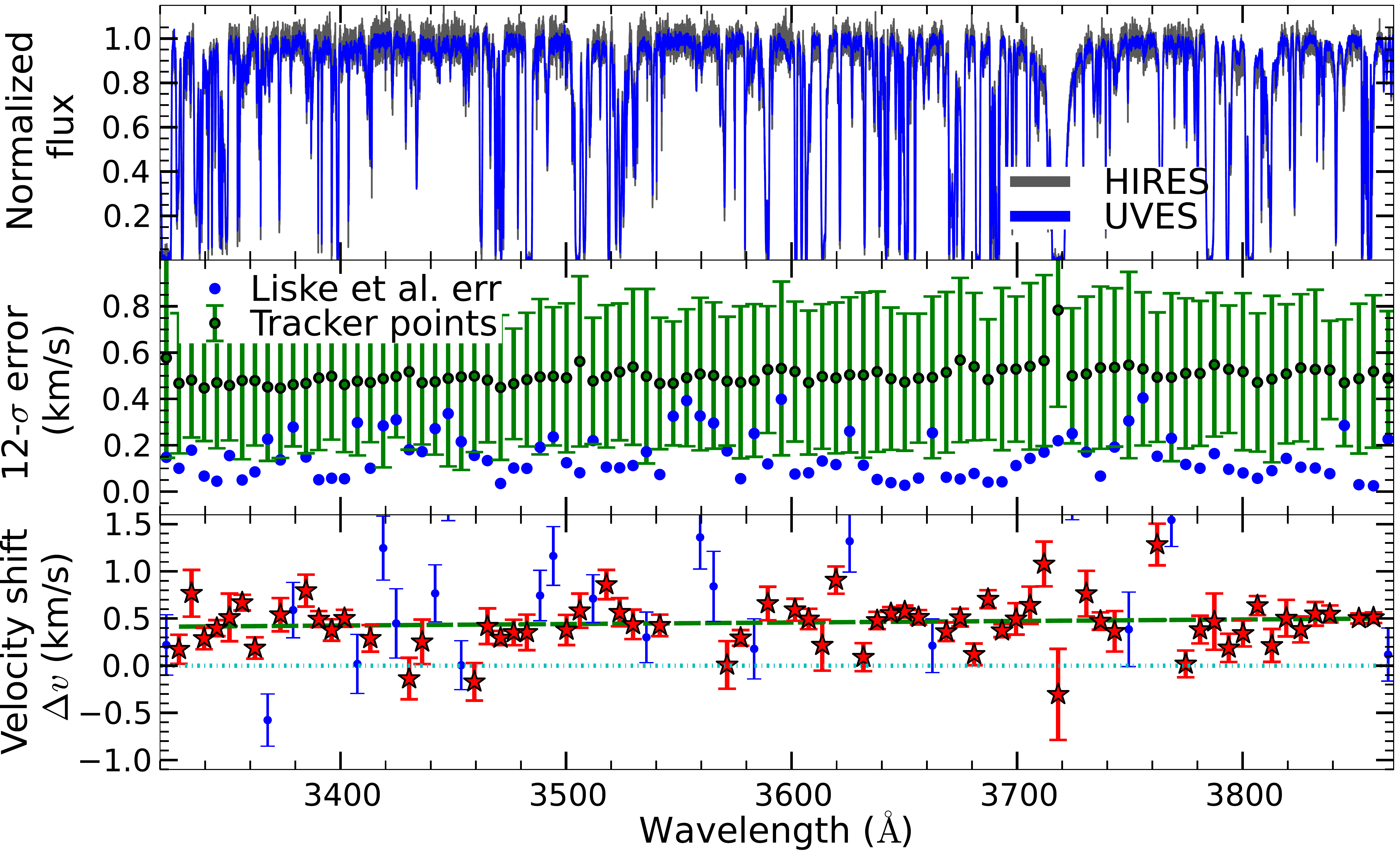}
\caption{\footnotesize Results from the DC method bluewards of the Lyman-$\alpha$ emission line for J2123$-$0050. The top panel shows the spectra from both UVES (blue line) and HIRES (grey line).  The middle panel shows the \citeauthor{Liske:2008:1192}~uncertainty (equation \ref{eq:Liske}) for each 500\,km\,s$^{-1}$-wide chunk and their corresponding tracker points. The error bars on the tracker points represent 12 times the RMS of the simulations. Chunks in which the Liske et al.~uncertainty is smaller than twelve times the RMS of the simulations below the tracker points are likely to provide reliable velocity shift estimates and uncertainties from the DC method; they are `selected' for further analysis.  The bottom panel shows the velocity shifts and their 1-$\sigma$ uncertainties derived from the DC method for all chunks.  The selected chunks are shown in red stars. Note the much smaller scatter in the velocity shift estimates for the selected points compared to the others.  The dashed green line is an weighted least squares fit to the selected chunks. }
\label{fig:blue_results}
\end{figure*}

In Figure~\ref{fig:blue_results} we present our results for the blue portion of the spectra.  The top panel covers all of the overlapping Lyman-$\alpha$ forest from the UVES and HIRES spectra, a wavelength range of 3071--3964\,\AA, a comparable range as covered in the red portion (Figure~\ref{fig:red_results}). The HIRES spectrum has a resolving power of $R\approx110000$ while the UVES spectrum has $R\approx53000$ in the blue, meaning that some of the sharper features, particularly metal lines, are less resolved in the UVES spectrum and consequently appear slightly broader and shallower than in the HIRES spectrum. This is less of a concern in the blue portion of the spectra than in the red (where it did not seem to have a substantial effect on the results) because the relatively broad Lyman-$\alpha$ forest lines are completely resolved in both spectra.  At 3420\,\AA\, the SNR for the HIRES spectrum is $\sim$25\,per 1.3\,km\,s$^{-1}$ pixel and for UVES it is $\sim$65\,per 1.5\,km\,s$^{-1}$ pixel. These SNR values, coupled with the broader but more numerous features than in the red portion, yield an average uncertainty in shifts calculated by the DC method of $\sim$0.06\,km\,s$^{-1}$. As with the red portion of the spectra, we want to use as large a chunk size as possible without causing the velocity shifts between the pair of spectra to be impacted by noise in the flux.  Firstly, there are more spectral features in the forest, so the chunks can be larger before they begin to be dominated by noise. Secondly, the many forest lines blend with each other and often produce spectral features that are much broader than metal-line complexes seen in the red part of the spectra. However, simple visual inspection reveals that $\sim$500\,km\,s$^{-1}$ chunks contain blends that are complex enough, with enough internal structure (e.g. sharp changes in flux), and almost always included some flatter, continuum-like regions, to reduce any risk of significant degeneracy between the parameters in the $\chi^2$ minimization, particularly the velocity shift and tilt parameters. Therefore, we chose a chunk size of 500\,km\,s$^{-1}$ bluewards of the Lyman-$\alpha$ quasar emission line. For higher (lower) redshift Lyman-$\alpha$ spectra, where the number-density of forest lines is larger (smaller) than in our spectra, one would expect to use larger (smaller) chunks.

The center panel of Figure~\ref{fig:blue_results} shows the \citeauthor{Liske:2008:1192}~velocity uncertainties and their corresponding tracker points.  However, the choice of $N$ for the tracker point selection process took into account several aspects of the Lyman-$\alpha$ forest region which differ significantly from the red portion.  Since the Lyman-$\alpha$ forest region contains many broad features (from atomic hydrogen) and also numerous narrow features (metals and, for this QSO, H$_2$ and HD) in every chunk, there is a much smaller dynamic range in the \citeauthor{Liske:2008:1192}~uncertainties. That is, the most reliable and least reliable chunks are less distinguished from each other. In principle, almost all chunks offer a reasonably reliable velocity shift estimate (and uncertainty). However, one may choose to only select the most reliable chunks, but this requires a slightly more careful tuning of the $N$ parameter.

The bottom panel of Figure~\ref{fig:blue_results} shows our results from the blue portion of J2123$-$0050.  Just as with the red portion of the spectra (Figure~\ref{fig:red_results}), the points identified as reliable show a much smaller scatter in velocity than those associated with regions of the spectra containing less spectral information. The reliable points show a general consistency in their velocity shift values, with a possible trend towards larger velocity shifts between the spectra at longer wavelengths. We find a slight offset between UVES and HIRES, similar to that found for the red portion, 0.44$\pm$0.03\,km\,s$^{-1}$. As described for the red results, we fitted a line to the velocity shifts found for the selected (i.e.~reliable) chunks and increased the uncertainties (by adding a constant value in quadrature to them) to obtain $\chi^2_{\nu} \approx 1$. This gives a slope of $(1.7 \pm 1.8)$\,m\,s$^{-1}$\,nm$^{-1}$.  This result is $< 1\sigma$ from zero and is consistent with no velocity distortion between the UVES and HIRES spectra in the Lyman-$\alpha$ forest.

It is interesting to consider the precision with which we have constrained a linear velocity distortion above (i.e.~1.8\,m\,s$^{-1}\,$nm$^{-1}$) and its implications for varying-$\mu$ analyses. This portion of the spectra (3071--3421\,\AA) contains the H$_2$ and HD transitions at $z_{\rm abs}=2.059$ which \citet{Malec:2010:1541} and \citet{Weerdenburg:2011:180802} used to measure $\Delta\mu/\mu$. Over this wavelength range, our precision would correspond to a velocity distortion of $\sim$63\,m\,s$^{-1}$.  For the Lyman H$_2$ lines over this wavelength range, the $K$ coefficients -- their sensitivity to variations in $\mu$ -- vary fairly smoothly from $-0.03$ for the reddest transitions to $+0.02$ for the bluest transitions\footnote{For our simple considerations here, we ignore the fact that the Werner H$_2$ transitions towards the blue end of this wavelength range have $K$ values opposite in sign to the neighboring Lyman lines. \citet{Malec:2010:1541} and \citet{Weerdenburg:2011:180802} have demonstrated that the Werner lines contribute little to constraining $\Delta\mu/\mu$ in the HIRES and VLT spectra studied here.}. Using equation~\ref{eq:mu}, we find that this maximum distortion would correspond to a $\Delta \mu/\mu$ measurement on UVES being more positive than its corresponding HIRES measurement by $4.2$\,ppm.

This maximum (1-$\sigma$) systematic error of 4\,ppm is similar to the statistical uncertainty on $\Delta\mu/\mu$ derived from the lower SNR of the two spectra, the HIRES spectrum, by \citet{Malec:2010:1541}. That is, the DC method, applied to the Lyman-$\alpha$ forest, utilizes enough spectral information to identify and measure systematic velocity distortions which are important for varying $\mu$ measurements. Secondly, the size of the possible systematic error in $\Delta\mu/\mu$ created by the marginally significant velocity distortion we find, is consistent with the very small (and insignificant) difference between the $\Delta\mu/\mu$ values found by Malec et al. (2010) and van Weerdenburg et al. (2010): $+5.6\pm5.5_{\rm stat}\pm2.9_{\rm sys}$ and $8.5\pm3.6_{\rm stat}\pm2.2_{\rm sys}$\,ppm respectively. That is, in this case, the DC method is consistent with the direct comparison of the $\Delta\mu/\mu$ values derived from the same pair of spectra. However, we emphasize that, by utilizing additional spectral information, not just the transitions used to derive $\Delta\mu/\mu$, the DC method enables the discovery of systematic errors to which a simple comparison of the $\Delta\mu/\mu$ values is not sensitive.

Recently, \citet{Rahmani:2013p1870} compared VLT/UVES asteroid (i.e.~reflected solar) spectra with a higher-resolution Fourier Transform spectrum of the Sun \citep{Kurucz:2005p1902,Kurucz:2006p1905} and identified velocity distortions over the wavelength range $\sim$3300--3900\,\AA\ with significant slopes, ranging between 2 and 6\,m\,s$^{-1}$\,nm. If distortions as strong as 6\,m\,s$^{-1}$\,nm were present between the HIRES and UVES spectra studied here, over the same wavelength range, they would have been detected by the DC method in Figure~\ref{fig:blue_results}. Nevertheless, it is worth noting that the sign of the (albeit statistically insignificant) distortion we identified, is the same as those identified by \citet{Rahmani:2013p1870} in the their UVES asteroid exposures (assuming that any distortions in the HIRES spectra are negligible).

\subsubsection{Intra-order distortions}

From the analyses of \citet{Griest:2010:158} and \citet{Whitmore:2010:89} we expect there to be some velocity distortions present on the scale of echelle orders (as opposed to the much longer-range distortions considered above).  \citet{Griest:2010:158} measured peak-to-peak intra-order distortions in individual HIRES quasar exposures ranging from 300\,m\,s$^{-1}$ to 800\,m\,s$^{-1}$ and \citet{Whitmore:2010:89} found peak-to-peak intra-order distortions in individual UVES exposures from 100\,m\,s$^{-1}$ to 200\,m\,s$^{-1}$.  Therefore, it is quite possible that pairs of spectra, each comprising many exposures, such as those of J2123$-$0050 considered here, could also contain these intra-order distortions. We search for evidence of intra-order distortions below using the DC method.

\citet{Griest:2010:158} and \citet{Whitmore:2010:89} used spectra of their objects observed through an iodine (I$_2$) cell to establish two different wavelength scales: that from the usual ThAr comparison lamp exposures and that from the dense I$_2$ forest of lines when compared to a laboratory Fourier Transform Spectrum. This allowed them to determine, to high precision, velocity shifts between the two wavelength scales as a function of position along each echelle order falling within the wavelength range covered by the I$_2$ transitions ($\sim$5000--6200\,\AA). By comparison, the DC method only utilizes the spectral information in the object spectra themselves, not the very dense information contained in ThAr or I$_2$ spectra. Therefore, searching for intra-order distortions within individual orders of our J2123$-$0050 spectra is not really possible; there is simply not enough spectral information, even in the Lyman-$\alpha$ forest, to provide enough reliable velocity shift measurements across any given echelle order.

\begin{figure*}[]\vspace{0.0em}
\epsscale{1.0}\plottwo{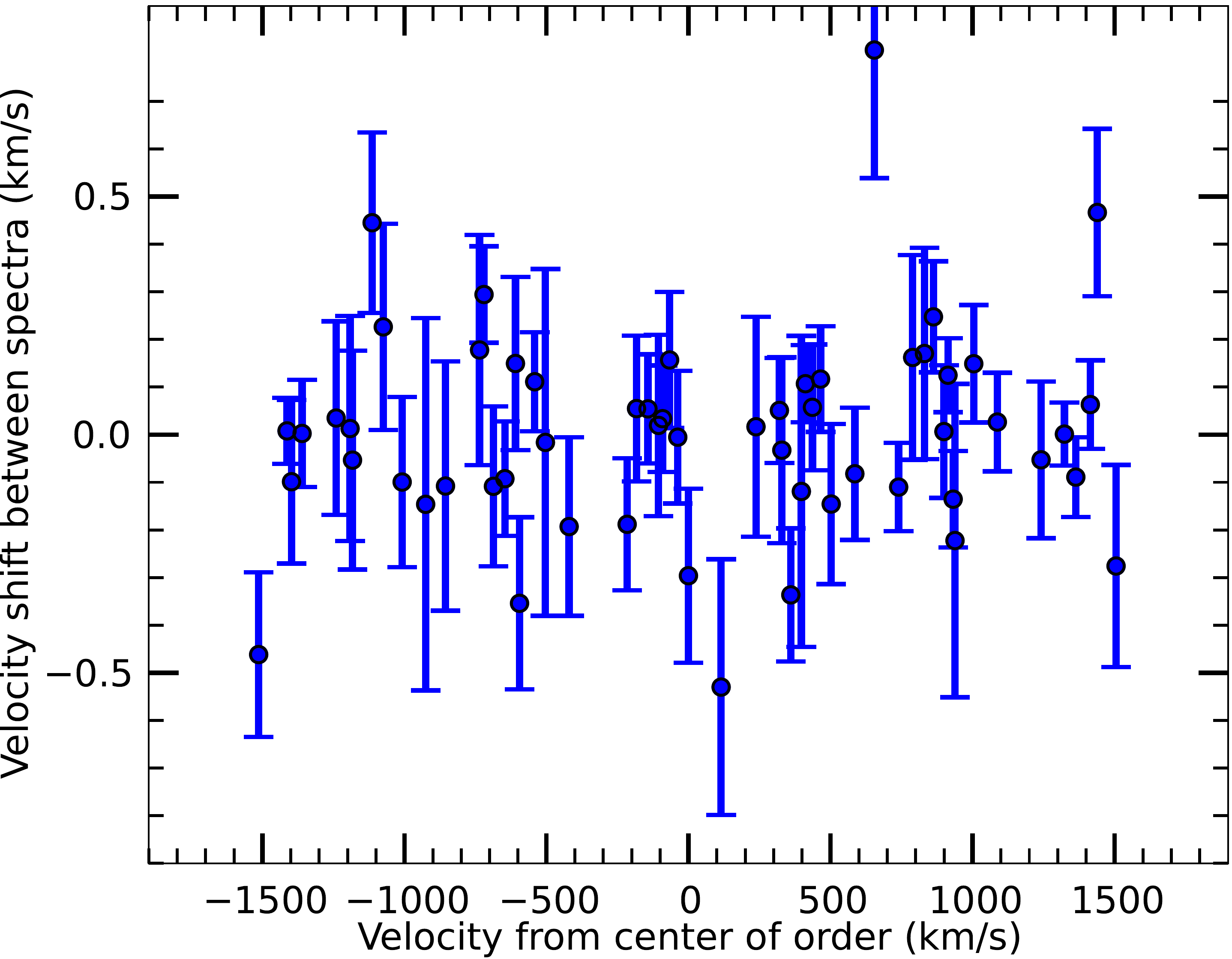}{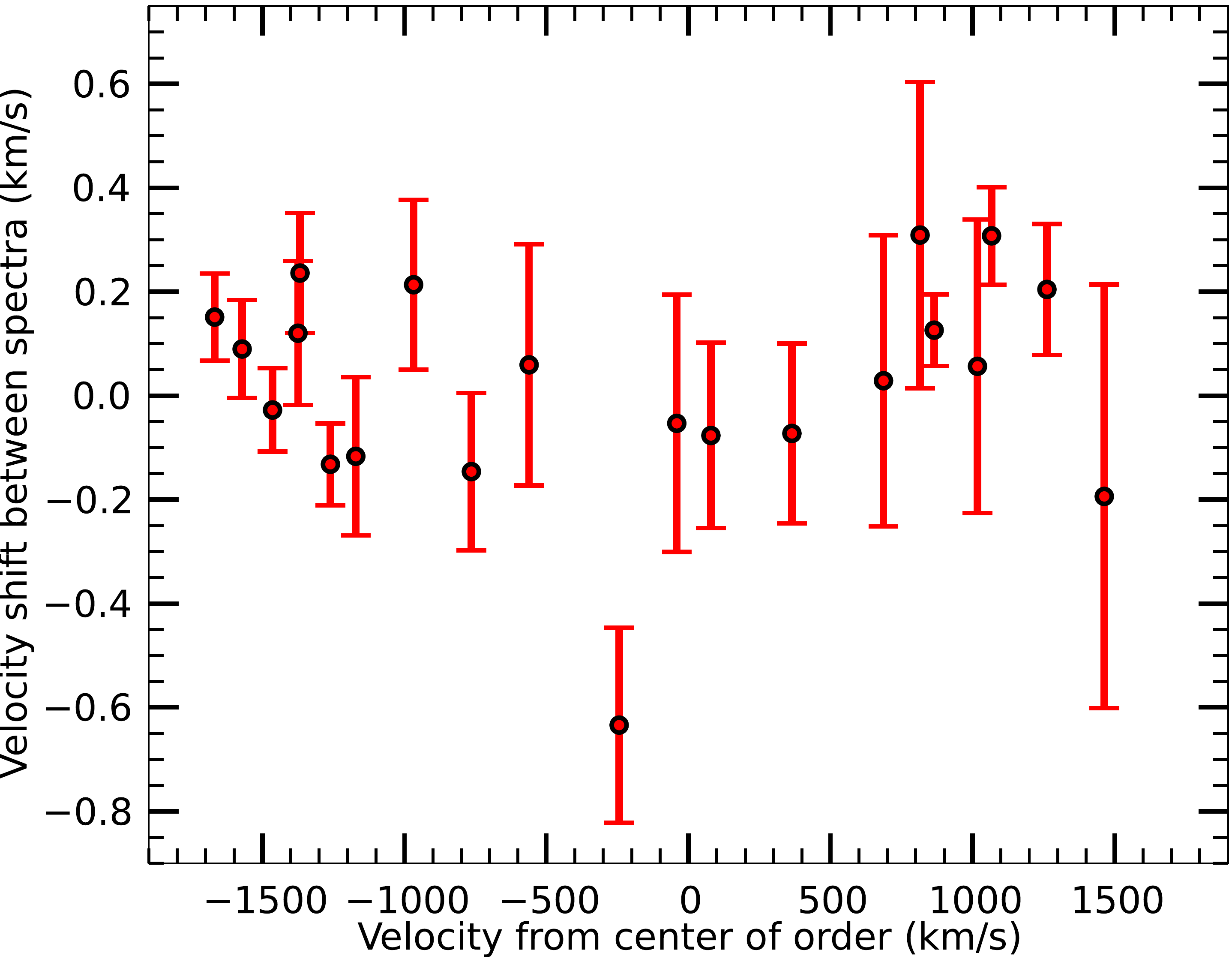}
\caption{\footnotesize Residuals from the lines of best fit in Figures~\ref{fig:red_results} \& \ref{fig:blue_results} relative to the position of the chunks from the centers of their respective Keck/HIRES echelle orders.  The left plot shows the results from the blue part of the spectra and the right plot shows the red portion.  No obvious evidence exists in either plot for intra-order distortions in HIRES which have the same general shape and amplitude in all echelle orders.}
\label{fig:HIRES_stack}
\end{figure*}

To remedy this lack of features we ``stack'' the velocity shift information for all echelle orders in each portion of the spectra (blue and red). To do this we take the residuals between the reliable $\Delta v$ values and the lines of best fit in Figures~\ref{fig:red_results} \& \ref{fig:blue_results}, and calculate how far away, in velocity, each chunk is from the center of its Keck/HIRES echelle order. This information is plotted in Figure~\ref{fig:HIRES_stack} for the blue and red portions of the spectra separately.  We find no obvious evidence in Figure~\ref{fig:HIRES_stack} for intra-order distortions which have a similar shape and amplitude in all the Keck echelle orders, in either the blue or red portions. While Figure~\ref{fig:HIRES_stack} assumes that the distortions are coherent among the echelle order structure of HIRES, we found similar results when stacking the velocity shift measurements with respect to their positions on the VLT/UVES orders.

There are several possibilities for why Figure~\ref{fig:HIRES_stack} shows no distortions but \citet{Griest:2010:158} and \citet{Whitmore:2010:89} find clear evidence for them. Firstly, these authors examined individual echelle orders in individual object exposures while SNR considerations force us to work with many exposures combined into a single spectrum.  It is possible that when the orders are combined to make the final spectrum, some of the distortions are dampened by the overlap of neighboring orders.  Another possibility arises from comparing spectra from different spectrographs. Because the order lengths are different in each spectrum, any coherent intra-order distortions across each spectrum will be differently phased and may substantially cancel each other out when analyzed with the DC method. On the other hand, they may reinforce each other, making the effect more pronounced.

\begin{figure}[]\vspace{0.0em}
\epsscale{1.0}\plotone{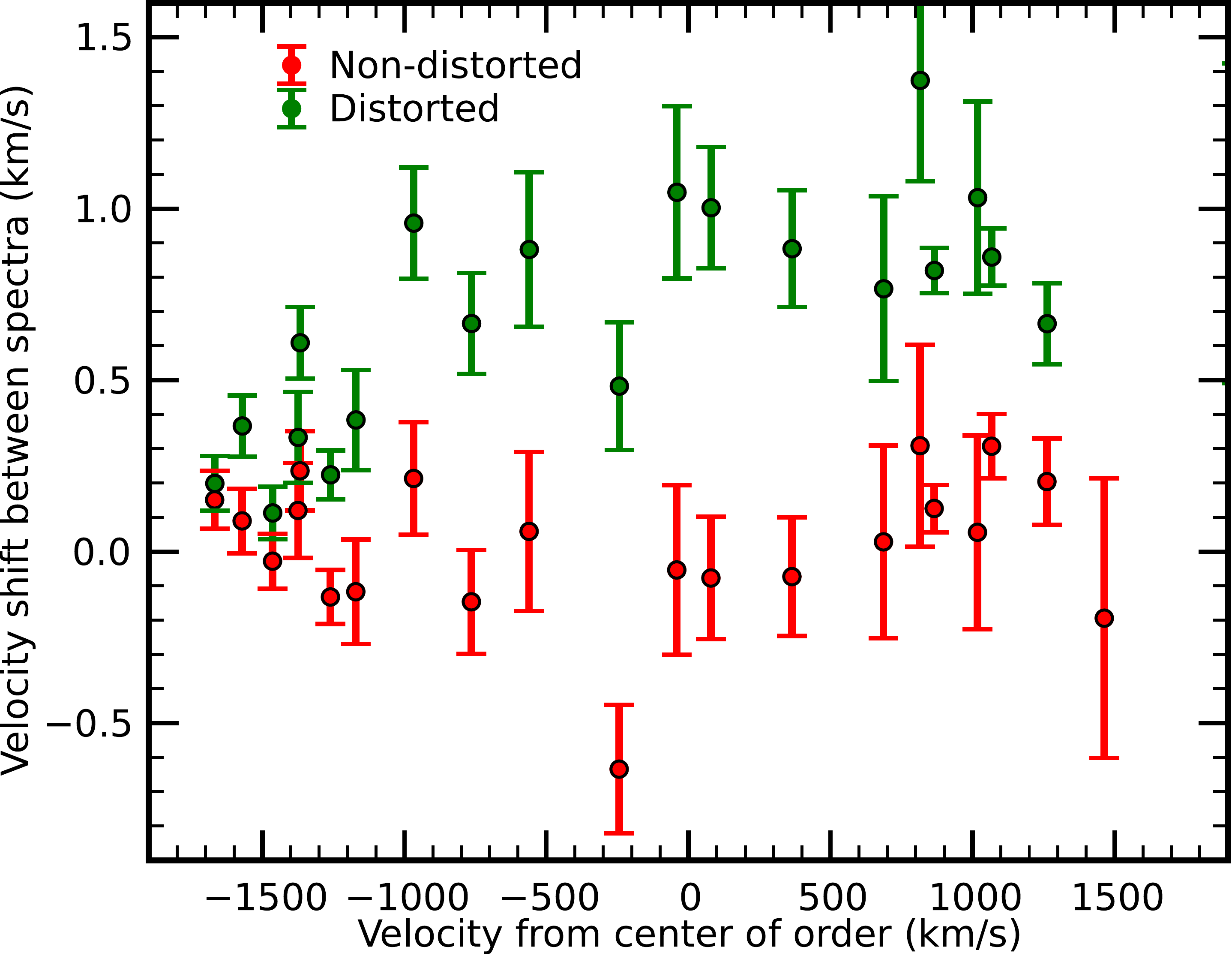}
\caption{\footnotesize Same as Figure~\ref{fig:HIRES_stack} but with an artificial intra-order distortion added to the red portion of the Keck/HIRES exposures.  The red points show the results from the original spectra while the green points are affected by the 800\,m\,s$^{-1}$ peak-to-peak saw-tooth distortion added to each HIRES echelle order. Note that, in the center of the orders, the 800\,m\,s$^{-1}$ artificial distortion is recovered while, towards the order edges, little evidence of it can be seen.}
\label{fig:order_distort}
\end{figure}

To test whether the DC method is sensitive to intra-order distortions in our spectra of J2123$-$0050, given the echelle order structures of HIRES and UVES, we introduced a saw-tooth distortion in all of the HIRES orders.  We took the extracted, wavelength-calibrated echelle orders before they were combined into a single spectrum and introduced a distortion with a peak-to-peak amplitude of 800\,m\,s$^{-1}$ into each echelle order.  This distortion causes the edges of the orders to have a negative velocity shift and the centers of the orders to have a positive velocity shift.  We then recombine the spectrum in the same manner as described by \citet{Malec:2010:1541} and reapply the DC method in the same fashion used to produce Figure~\ref{fig:HIRES_stack}. Figure~\ref{fig:order_distort} shows that we recover the additional 800\,m\,s$^{-1}$ velocity distortion in the center of the HIRES orders. However, we do not recover any distortions at the edges of orders.  The regions that show little sign of distortion at velocities (relative to the order centers) where we expect neighboring orders to overlap. There, the saw-tooth nature of the artificial distortion causes overlapping orders to substantially cancel out the distortions. Therefore, if real intra-order distortions do exist in the HIRES spectra, but not the UVES spectra, we would expect to see them in Figure~\ref{fig:HIRES_stack} after applying the DC method to the combined, multiple-exposure Keck and VLT spectra of J2123$-$0050.

Figure~\ref{fig:order_distort} demonstrates that the DC method can detect intra-order distortions, allowing us to set a rough upper limit to any intra-order distortions between HIRES and UVES which have a similar shape and amplitude across all orders.  From Figure~\ref{fig:order_distort}, it is clear that such intra-order distortions with a peak-to-peak amplitude $\ga$500\,m\,s$^{-1}$ would be detectable, so we treat that as a conservative upper limit.  Since we expect overlap of neighboring echelle orders to diminish the effects of intra-order distortions, it follows that we would see less evidence for intra-order distortions in the blue end of the spectrum, compared to the red, because the overlap is much greater for bluer orders (both UVES and HIRES are grating cross-dispersed spectrographs). This implies that there should be little impact on the measurements of $\Delta \mu/\mu$, because such measurements usually utilize the bluest set of echelle orders to access the Lyman and Werner H$_2$ transitions.  However, it is important to note that, because the DC method can only detect distortions \textit{between} pairs of spectra, it is possible that intra-order distortions with the same (or similar) shape and amplitude in both HIRES and UVES would not be detected. Having said that, if the two telescopes have similar distortions, then similar samples of absorbers on both telescopes, from which (spurious) measurements of $\Delta \mu/\mu$ or $\Delta \alpha/\alpha$ have been made, would be affected in the same manner, and this would not lead to different average measurements from the two different telescopes (like, for example, those leading to evidence for a variation in $\alpha$ across the sky).

\begin{figure*}[]\vspace{0.0em}
\epsscale{1.0}\plotone{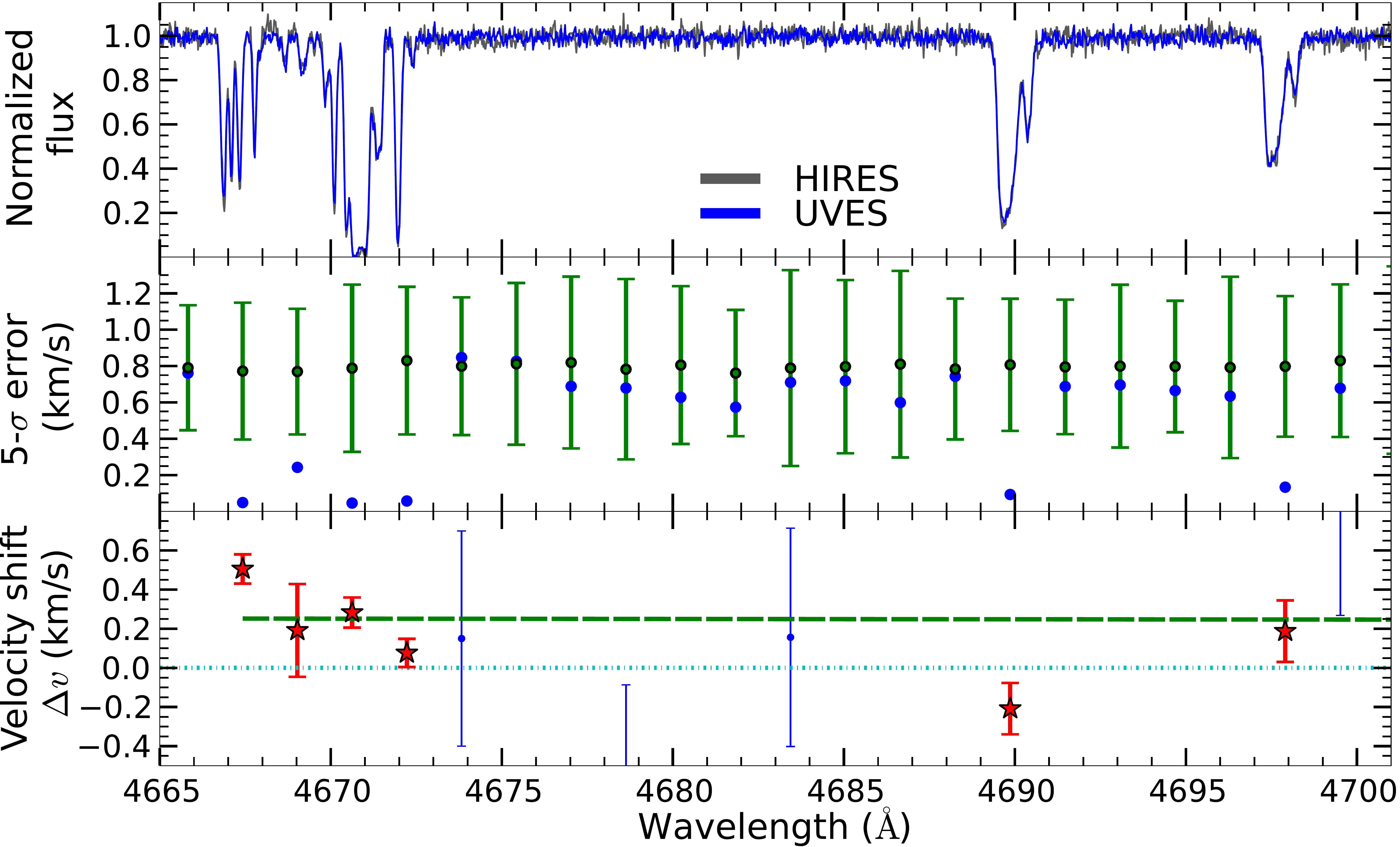}
\caption{\footnotesize A possible detection of a coherent, intra-order distortion.  The color-coding of lines and symbols is the same as in Figs.~\ref{fig:red_results} \& \ref{fig:blue_results} but here a smaller chunk size of 100\,km\,s$^{-1}$ is used to explore the possible distortion in more detail. The top panel shows the spectra of the region possibly affected by a short-range distortion.  The bottom panel shows velocity shifts for the selected chunks (those with sufficient spectral information) and the corresponding line of best fit.}
\label{fig:short_distort}
\end{figure*}

Despite not finding any overall evidence for intra-order distortions, we do see one example of a possible coherent, short-range distortion.  At the bluest edge of the red portion of spectra shown in Figure~\ref{fig:red_results}, from 4665\,\AA\, to 4700\,\AA, we noticed a negative trend in velocity shift between the HIRES and UVES spectra.  We have further explored this region in Figure~\ref{fig:short_distort} by using a smaller chunk size (100\,km\,s$^{-1}$) than in Figure~\ref{fig:red_results} (200\,km\,s$^{-1}$). This $\sim$30\,\AA\, region has a almost 500-m\,s$^{-1}$ distortion across it, including some of the only negative values of velocity shift (HIRES blueshifted relative to UVES) measured in the whole spectrum. We fitted a weighted linear regression to the velocity shifts of the selected features in the bottom panel of Figure~\ref{fig:short_distort}.  Despite being able to see clear, coherent distortions by eye, it is obvious that a linear fit does not describe these distortions well.  It is unclear what causes this effect. Being in the red portion of the spectra, this would not have had any effect on the measurements of $\Delta \mu/\mu$ performed by \citet{Malec:2010:1541} and \citet{Weerdenburg:2011:180802}.  However, because only a 350-\AA\ range of these spectra contain H$_2$ and HD which were used to constrain $\Delta\mu/\mu$, short range distortions like the one seen at 4665--4700\,\AA\ could have a marked effect on $\Delta\mu/\mu$ if present in the blue portion of one or both spectra.

\section{Conclusions}
\label{sec:DC_conclusions}

We have demonstrated in this paper that the Direct Comparison (DC) method is a robust tool that allows us to measure velocity shifts between pairs of spectra accurately and with a reliable uncertainty estimate. It improves on previous methods of detecting wavelength-dependent velocity shifts (i.e.~velocity distortions) between spectra. Previous works have typically used line-fitting or cross-correlation methods (e.g.~\citealt{King:2012:3370} and \citealt{Kanekar:2011:L12} respectively). While those methods have established the importance of searching for velocity distortions in pairs of spectra for varying-constant studies, they have important disadvantages. The line-fitting method relies on fitting previously-identified spectral features, rendering it model-dependent, labor-intensive, and time-consuming. While the cross-correlation method is model-independent, it cannot reliably recover sub-pixel velocity shifts or robust uncertainties without detailed Monte Carlo simulations. The DC method is a model-independent approach which does not require prior identification of spectral features, making it fast and straight-forward to apply to large samples of (pairs of) spectra. While our Monte Carlo testing of the DC method shows that it underestimates the uncertainty at ${\rm SNRs}\lesssim7$\,per pixel, neither line-fitting or cross-correlation methods can perform reliably at these low SNRs either.

We applied the DC method to two spectra of J2123$-$0050 that have previously been used to measure $\Delta\mu/\mu$: \citet{Malec:2010:1541} on Keck and \citet{Weerdenburg:2011:180802} on the VLT. While these previous works naturally focused on the H$_2$/HD lines in the Lyman-$\alpha$ forest, we also compared the spectra redwards of the Lyman-$\alpha$ emission line to more comprehensively test for systematic errors between Keck and the VLT. We found no strong evidence for velocity distortions between these two spectra over relatively large wavelength ranges. For the red portion, which contained several complex metal absorption systems over a wavelength range of $\sim$4650--5700\,\AA, we found that the HIRES spectra are blueshifted relative to the UVES spectra by $0.31\pm0.05$\,km\,s$^{-1}$. However, the change in this offset, i.e.~the velocity distortion, across this wavelength range is consistent with zero: $(0.4\pm1.7)$ m\,s$^{-1}$nm$^{-1}$ -- see Figure~\ref{fig:red_results}. This amounts to a maximum (1-$\sigma$) velocity distortion of 42\,m\,s$^{-1}$ over a 1050\,\AA\ range. Over a shorter wavelength range of $\sim$4650--5100\,\AA\ we found a non-zero distortion of $(4.9\pm2.2)$ m\,s$^{-1}$nm$^{-1}$.  One possible explanation for this difference is that a simple linear fit may not actually be a good representation of the distortions present in the spectra. We found a similar velocity offset, $0.44\pm0.03$\,km\,s$^{-1}$, for the blue portion of the spectra, 3071--3421\,\AA, with a similar, null detection of $(1.7\pm1.8)$ m\,s$^{-1}$nm$^{-1}$ -- see Figure~\ref{fig:blue_results}. The precision of this constraint would have been sufficient to detect the larger velocity distortions in the same wavelength range identified by \citet{Rahmani:2013p1870} in UVES asteroid spectra.

We also examined the possibility of velocity distortions between the Keck and VLT spectra on shorter, sub-echelle order scales, as we might expect from the results of \citet{Griest:2010:158} and \citet{Whitmore:2010:89}. These works revealed evidence for such `intra-order distortions' in individual quasar exposures, with peak-to-peak amplitudes between 300 and 800\,m\,s$^{-1}$ for HIRES and 100 and 200\,m\,s$^{-1}$ for UVES. However, when applying the DC method to our spectra, each of which comprises many exposures, we found no evidence for intra-order distortions with coherent shape and amplitude across all orders. By inserting artificial intra-order distortions into the individual Keck exposures, we found that our application of the DC method would have recovered them, particularly near the order centers, if their peak-to-peak amplitude exceeded $\sim$500\,m\,s$^{-1}$. The overlap of neighboring echelle orders (in wavelength space) substantially diminished the sensitivity to these artificial intra-order distortions near the order edges. Distortions with peak-to-peak amplitudes $\gtrsim$500\,m\,s$^{-1}$ might have evaded detection if they had a similar shape and amplitude in the Keck and VLT spectra. That is, because the DC method only detects distortions \textit{between} spectra, similar distortions in both spectra may not be detected with the DC method. It is important to realize that, even if this were the case, it would not cause different values of $\Delta\alpha/\alpha$ or $\Delta\mu/\mu$ to be found from Keck and VLT.

One of the main motivations for the DC method was to develop a tool to check the recent \citet{Webb:2011:191101} and \citet{King:2012:3370} result concerning dipolar variation in $\alpha$.  The majority of their northern targets were measured on Keck/HIRES and give an overall negative value of $\Delta \alpha /\alpha$, while their southern targets, typically observed on the VLT/UVES, give an overall positive $\Delta \alpha /\alpha$.  The most obvious check on this result is to make sure that Keck and the VLT agree with each other and try to eliminate/understand all systematic errors.  The DC method allows us to map out any systematic distortions or velocity shifts between pairs of spectra from the two telescopes without having to know what causes them. Though we found no statistical support for distortions in one particular pair of Keck and VLT spectra, a distortion of the sign and magnitude we measured, \emph{if} found to be statistically significant, would imply that the $\alpha$-dipole has a larger amplitude than currently measured. In addition to comparing the VLT and Keck telescopes, our future work will compare spectra from UVES and HIRES to spectra from the Subaru Telescope's High Dispersion Spectrograph (HDS). Inclusion of a second northern telescope will help better establish whether or not the $\alpha$-dipole is the result of systematics between telescopes.

We have begun applying the DC method to new observations of $\Delta \alpha/\alpha$. While this paper focussed on using the DC method to compare spectra from different telescopes, we are also applying it to compare different exposures of the same object taken with the same telescope. Molaro et al.~(in prep) present the first measurements of $\Delta \alpha/\alpha$ from a new observational program using the VLT/UVES. They present high-SNR observations of a single quasar at high resolution ($R\sim60000$), observed in many individual exposures with ${\rm SNR}\sim20$ per exposure. Because VLT/UVES is a grating cross-dispersed slit spectrograph, deviations in the position of the object (quasar) in its slit effectively impart velocity shifts to the recorded spectrum. Therefore, exposures of the same object, taken at different times, will have small velocity shifts between them. To detect and remove these velocity shifts between exposures, Molaro et al.~applied the DC method to each exposure prior to combining them into a final spectrum for measurement of $\Delta \alpha/\alpha$. Future papers will examine a larger sample of quasars from the Molaro et al.~VLT program in conjunction with spectra of the same quasars from Keck/HIRES. The DC method will be applied to these spectra, both to correct velocity shifts between exposures and to check for systematic velocity distortions between UVES and HIRES.

This technique may also find application in other fields of astronomy requiring precise spectroscopy. For example, the DC method could be used to determine radial velocity changes in stellar spectra. Current techniques for identifying extra-solar planets match stellar spectra with templates to look for velocity shifts caused by an additional orbiting body \citep{Butler:1996:500}. The DC method can be used to accurately and robustly determine the velocity shift between the template and individual stellar exposures. With spectra observed with highly stable instruments, like the HARPS spectrograph \citep{Mayor:2003:20}, it may be possible to simply compare stellar exposures directly to each other, or to their average, with the DC method, to extract the radial velocity variations.

Another interesting application of this method could be in measuring cosmological drift, as first proposed by \citet{Sandage:1962:319}. To perform the `Sandage test', the redshift of Lyman-$\alpha$ forest absorption features \citep{Loeb:1998:L111} or radio 21-cm absorption \citep{Darling:2012:L26} could be compared at different epochs.  When Sandage first proposed this test, the instrumentation was not sensitive enough to detect the expansion of the universe on time-scales under $10^7$ years \citep{Sandage:1962:319}. \citet{Liske:2008:1192} calculated that with a 40 meter optical telescope it would be possible to detect cosmological shifts with observational epochs separated by $\approx 20$ years. Even smaller durations may be possible using 21-cm absorbers observed with the Square Kilometer Array \citep{Darling:2012:L26}.  By measuring spectral shifts over time, the expansion history of the universe could be obtained directly.  Firstly, an important consistency check must be applied to the first-epoch spectra -- there must not be significant systematic shifts or distortions between them -- before a second epoch of observations is considered. The DC method could provide a model-independent, automatic and robust method of testing this. Finally, after the second-epoch of spectra are observed, the DC method could be applied to robustly measure the drifts in the absorption features resulting from cosmic expansion.

\acknowledgments 
We thank Jonathan Whitmore and Adrian Malec for helpful discussions. This work is based on observations carried out at the European Southern Observatory under program No.~81.A- 0242 (PI Ubachs), with the UVES spectrograph installed at the Kueyen UT2 on Cerro Paranal, Chile. We thank J.~Xavier~Prochaska and Sara Ellison for providing the Keck spectrum of J2123$-$0050. Some of the data presented herein were obtained at the W.~M.~Keck Observatory, which is operated as a scientific partnership among the California Institute of Technology, the University of California and the National Aeronautics and Space Administration. The Observatory was made possible by the generous financial support of the W.~M.~Keck Foundation. The authors wish to recognize and acknowledge the very significant cultural role and reverence that the summit of Mauna Kea has always had within the indigenous Hawaiian community. We are most fortunate to have the opportunity to conduct observations from this mountain.
We thank the anonymous referee for their measured and insightful review and for helping us improve the manuscript.
MTM thanks the Australian Research Council for funding under the Discovery Projects scheme (DP110100866).





{\it Facilities:} \facility{Keck I (HIRES)}, \facility{VLT (UVES)}.

\end{document}